# The Where and How of Touch: A Review of Tactile Localization Research

Xaver Fuchs[1,2*], Jason A.M. Khoury[3*], Sergiu Tcaci Popescu[3], Tobias Heed[1,2#], Matej Hoffmann[3#]

[1]Department of Psychology, University of Salzburg, Salzburg, Austria

[2]Centre for Cognitive Neuroscience, University of Salzburg, Salzburg, Austria

[3]Department of Cybernetics, Faculty of Electrical Engineering, Czech Technical University, Prague, Czech Republic

[*#]equal contributions

## Abstract

Tactile localization is the seemingly simple ability to “tell” where a touch has occurred. However, how this ability is assessed, and what conclusions are drawn from experiments, depends on the theoretical ideas that inspire the research. Here, we review both theoretical frameworks and methodological approaches based on a systematic web-based literature search on tactile localization. After presenting current theories of tactile localization, we discuss task characteristics that differentiate current methodology for tactile localization into at least eight distinct types of experimental tasks. We describe these tasks, discuss their, often implicit, underlying assumptions and cognitive requirements, and relate them to the theoretical approaches. We then compare, in an exemplary manner, the tactile localization results reported by a subset of studies and demonstrate how some methods are associated with specific biases, illustrating that the choice of experimental method significantly affects the conclusions drawn from the results. Our review suggests that the field currently lacks a clear concept of the specific processes induced by the various experimental tasks and, thus, calls for concerted efforts to clarify and unify currently diverse, fragmented, and partly inconsistent theoretical underpinnings of tactile spatial processing, flanked by dedicated data sharing to allow across-study analysis.

# 1 Introduction

The skin encloses our entire body and is, thus, our largest sensory organ. It is endowed with thousands of receptors for touch, temperature, and potentially harmful events that give rise to the experience of pain (Corniani & Saal, 2020). As the border between our body and the world, the skin senses contact with external objects both when objects touch us from the outside, and when we actively explore and act upon our environment. This makes the sense of touch also an essential component of the sensorimotor system, necessary for skillful action, as anyone can attest to who has attempted to zip a jacket while wearing gloves.

Psychology and neuroscience textbooks often introduce the tactile modality by describing some basic aspects of touch, for example, how stimulus intensity and duration are neurally encoded in the firing rates of fast and slowly adapting fibers, and how spatial discrimination ability is related to the density of tactile receptive fields (e.g., Birbaumer & Schmidt, 2010; Kandel et al., 2021). A related, yet more intricate aspect that textbooks rarely describe in detail is how we infer a touched *location* on our body and perform actions like reaching for that touched location. This ability is typically termed *tactile localization*.

Tactile localization is an adaptive behavior, which can be easily appreciated with the simple (and frequently invoked) example of a mosquito landing on your body: it is necessary to know where it sits to swipe it away. Our body can be configured very flexibly into various postures; therefore, there is no simple one-to-one relationship between a given location on the skin and where that location is in (Cartesian) space. Rather, posture must somehow be integrated to make a sensible motor response to a tactile stimulus. Tactile localization is currently an umbrella term for both aspects: where a touch is perceived on the skin, and where it is perceived in space.

## 1.1 Tactile localization on the skin

The first aspect, localizing touch on the skin, is referred to by various terms, e.g. 'somatic', 'somatotopic', 'skin-based', and 'anatomical' (e.g., Heed et al., 2015; Longo et al., 2010; Medina & Coslett, 2010). It is typically conceptualized as a 'purely tactile' process, meaning that it does not require information retrieved from other sensory modalities or stored representations beyond those that relate to the skin itself. As an example, imagine a health check, in which the doctor asks you to close your eyes, touches individual fingers, and asks you to call out which finger was touched. In principle, this task requires no more than assigning the stimulus to a region of the skin, setting aside, for now, the transformation into a semantic code and a verbal response. Localization on the skin is often considered a first processing step (e.g., Longo et al., 2010; Medina & Coslett, 2010) and is closely tied to basic attributes of neural processing in the periphery and in early somatosensory areas of the brain, such as the shape and size of receptive fields. A touch sensation on the skin is produced by a physical stimulus that excites cutaneous receptors for pressure, vibration, or skin stretch. Neural signals travel through the spinal cord to the brain stem, thalamus, and primary somatosensory cortex (S1) in the postcentral gyrus. Within S1, the spatial layout of neurons resembles the layout of the skin: neighboring skin receptors are routed to neighboring neurons in S1, a feature termed somatotopic organization. Famously, the cortical 'map' is highly distorted, with the relative sizes of the cortical areas reflecting the number of tactile fibers of their source regions on the body rather than those regions' 3D size. This is why the mouth and the fingers cover an over-proportionally large section of the cortical surface in S1. This organization was initially identified with electrical brain stimulation during brain surgery (Penfield, 1958) and is often depicted by Penfield and Boldrey's (1937) somatosensory 'homunculus', a large-handed mannequin, whose disproportional body parts illustrate the surface of the corresponding cortical sections in S1 (see Fig. 1 for a version in which the mannequin's body regions are drawn over the respective matched S1 regions).

These findings raise two much-discussed points. The first is that we don't perceive our body to look or feel in the distorted dimensions of the homunculus. Thus, there is a marked discrepancy between the homuncular neural organization and both our behavior and our conscious experience. We'll return to this point later (see section 1.4 Body representations). The second point is the question of whether and how the homuncular organization of neurons can be turned into information accessible to the brain. Naively, one could think that the brain just needs to 'look up' which S1 neurons are firing to know where the touch was on the body. However, there is no 'little man' in the brain who observes or 'looks up' which neurons are firing and then 'tells' another structure where the touch occurred (O'Regan, 2010; see chapter 13) (see Fig. 1).

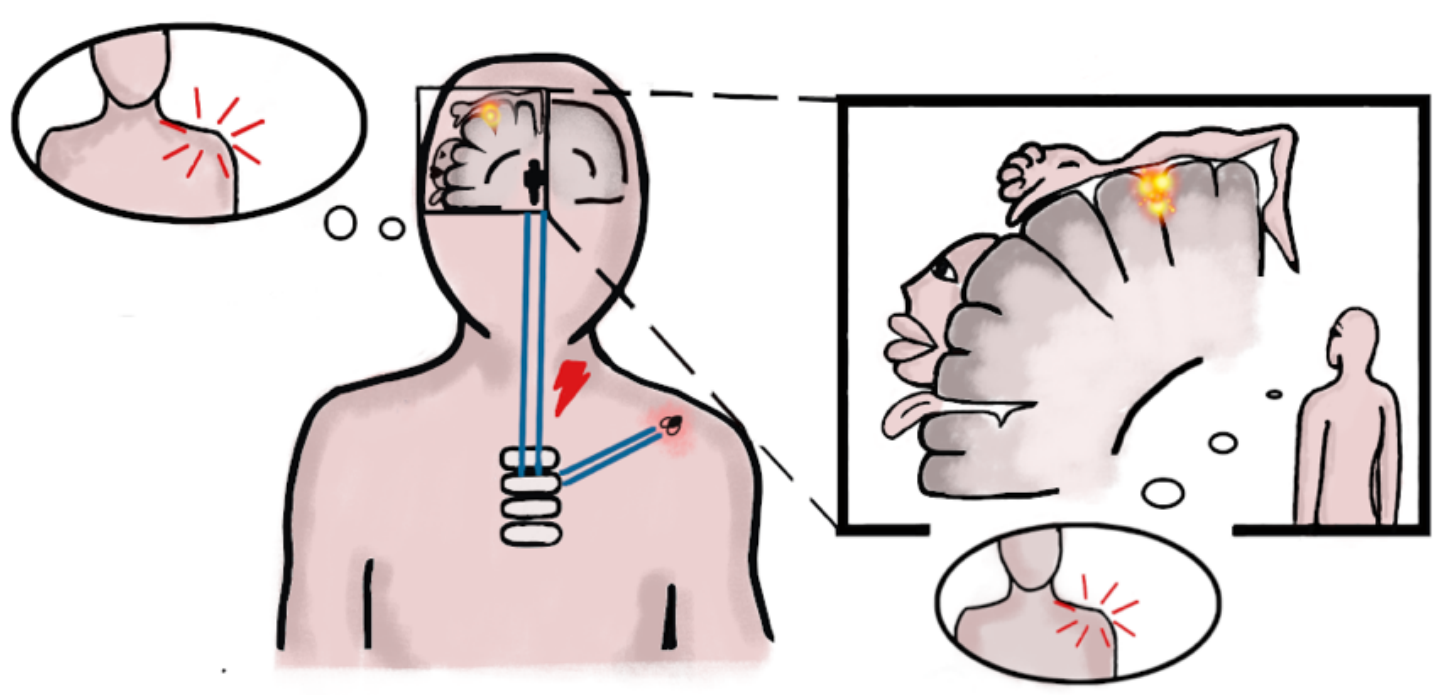

Fig. 1: Illustration of the concept that neural activity within the somatotopic map of the primary somatosensory cortex (S1) 'solves' the problem of tactile localization. A fly landing on the shoulder causes activation of the shoulder's representational S1 area. But the necessity of a 'little man in the brain' looking at the activation merely shifts the problem into the 'little man's' brain.

We must avoid the trap of equating the existence of the S1 map with an explanation for tactile localization; yet, implementations for reading out map-like content from neural population codes exist (e.g., Seung & Sompolinsky, 1993). Presumably, the cortical map, therefore, reflects a purposeful or energy-efficient neural architecture for tactile location coding, for instance by allowing efficient connections of nearby neurons (and, thus, skin locations) to detect correlated stimulation patterns when a larger skin area is touched (cf Seizova-Cajic et al., 2020), or to delineate body parts based on de-correlated tactile input – akin to the organization of cortex into columns known from primary visual cortex and motor cortex (e.g., Graziano & Aflalo, 2007; Mountcastle, 1997; Yoshimura et al., 2005). Moreover, the map-like structure of S1 likely plays a role in organizing larger-scale networks encompassing the thalamus, S1, and other parietal regions (Ghazanfar et al., 2000; Nicolelis et al., 1998).

## 1.2 Tactile localization in (external) space

When the skin location of a tactile stimulus is known, it can be further processed to derive the 3D location of the touch in space. It is currently a prevalent idea that any kind of movement requires a 3D coordinate of the goal location, that is here, the touch. Although it is difficult to trace the origin of this idea, it is tightly connected to sensorimotor research involving visual stimuli, which are natively in a 3D space, and reports of sensorimotor processes using a common, eye-centered (i.e., vision-related) spatial code (e.g., Andersen & Cui, 2009; Batista et al., 1999). This kind of function is referred to as *spatial localization of touch*(e.g., Longo et al., 2010) or *tactile remapping* (e.g., Heed et al., 2015; Shore et al., 2002). Note, our use of the term “3D” refers merely to the fact that tactile information can be related to locations in external space but is not meant to imply that the brain necessarily uses a Cartesian 3D coordinate system as we use it in geometry.

It is clear that humans can derive a spatial location from a touch. For instance, when a person holds out her arm, is then touched on the hand, retracts the touched arm, and is subsequently asked to point to where the touch occurred in space, a 3D goal location must be derived because the hand can no longer serve as a direct marker of where the touch had occurred. However, alternatives to such a 3D-spatial code exist for many other situations. When a person is touched on the left arm and wants to reach that location with her right hand, she could re-instate a body configuration from memory that reproduces the proprioceptive signals involved in the goal state (i.e., the right hand touching the left arm) without ever deriving a 3D-spatial location (e.g., Marcel et al., 2022). An analogue argument can be made for eye movements, with the goal state being a head direction and eyes-in-head direction relative to the touched arm. Thus, though it is often asserted that a 3D location is required, this is not necessarily true. Similarly, theories of tactile localization have included tactile remapping as a potential or mandatory step during tactile processing, and we will return to this point below (see Section 2.3 Tactile Remapping).

## 1.3 Tactile localization and 'body representations'

We've established that the term tactile localization can pertain to a location on the skin and to a location in 3D space. So far, however, it is not clear how our cognitive system gets from one to the other. Moreover, recall from above the curious fact that we perceive the relative sizes of our body parts quite differently than what we might expect from the distorted sizes evident in the S1 homunculus. Finally, we are also able to perform actions that cannot be based on current sensory input alone, such as reaching for an itch on our nose or telling where our hand is in space even with our eyes closed. Such behaviors can implicate knowledge about spatial attributes of the body for which no direct sensory system exists, for example, the size and shape of body segments. Psychology and neuroscience often frame such knowledge as being organized in so-called body representations, that is, stored information about '*what the body is usually like*' (Carruthers, 2008). Different types of knowledge are posited to be stored in individual representations, so that current models typically propose multiple representations. To return to the homuncular vs. felt body part sizes, one idea has been that there must be a representation beyond that of the S1 homunculus in which the homuncular distortions are corrected for (Longo et al., 2010; Longo & Haggard, 2011).

Originally, the notion of multiple body representations for tactile localization stems from observations of neurological patients. For example, Paillard (1999) described a patient who could not indicate where he had been touched on a drawing of the body but was unimpaired in his ability to point to the location on the body proper. Vice versa, another patient could not point to touched locations on the arm with closed eyes but was unimpaired in showing them on a drawing. These and other double dissociations (e.g., Anema et al., 2009; Liu et al., 2020) are the basis of the influential, albeit controversial, idea of two parallel processing routes in body perception: a perceptually and conceptually driven system, often referred to as the *body image*, and an action-based and more implicit route, referred to as the *body schema* (Head & Holmes, 1911; for overviews see Gallagher, 2005; Ataria et al., 2021; Poeck & Orgass, 1971). These concepts have been further subdivided in later models (e.g., Longo et al., 2010; Longo & Haggard, 2011) that we will discuss in more detail below (see Section 2.1 Representational models).

## 1.4 From theories to experiments: aim and scope of our review

It seems straightforward to expect that theoretical concepts, such as localization on the skin, localization in space, and various body representations, are each probed with tailored experimental tasks. Research, however, has often not drawn clear distinctions of theoretical underpinnings, and many experimental paradigms cannot easily be assigned to one or another category. Similarly, the interpretation of experimental results rests, of course, on the theoretical ideas that the respective researcher intends to put into the experiment – but many theoretical aspects are often overlooked or ignored. As a result, what is pooled under the umbrella term of tactile localization is likely not a single, well-defined mechanism, but a set of behaviors tailored to a multitude of specific task demands. To understand which kind of cognitive strategies a method might allow for, which cognitive processes it may imply, and, consequently, which kind of underlying mechanism it assesses, we must closely inspect each task in the light of the different theories.

To aid this endeavor, we provide 3 building blocks derived from a systematic review of the tactile localization literature. In Part 1, we give a brief overview of influential theoretical concepts, or frameworks, of tactile localization. In Part 2, we evaluate, categorize, and examine the tasks employed in the reviewed literature, identify explicit and implicit assumptions about the underlying sensorimotor or cognitive processes, and (where possible) relate methods and frameworks to one another.

In Part 3, we showcase some quantitative comparisons of localization performance for several tasks and illustrate how specific methods produce distinct result patterns that may, in turn, bias interpretations and model building.

Overall, our review uncovered a remarkable diversity of experimental approaches. This heterogeneity reflects that tactile localization is highly context-dependent and, accordingly, evokes many different (sub-)processes depending on what participants are asked to do.

# 2 Theoretical frameworks of tactile localization

Theorizing about body representations dates back to the turn of the 19$^{th}$ and 20$^{th}$ century (e.g., Poeck & Orgass, 1971; Kalckert, 2021). In fact, Pick wrote in 1922: "The abundance of speculation and theoretical constructs in the doctrine of the loss of awareness of the body, i.e. of defects in the 'somatopsyche,' is inversely proportional to factual observations" (cited from Poeck & Orgass, 1971). In contrast, the current landscape appears reversed. As de Vignemont (2018, page 2) noted, "These last twenty years have seen an explosion of experimental work on body representations, which should help us shape and refine our theory of body awareness." Indeed, our review encompasses more than 100 articles with experiments on tactile localization alone. Yet, despite this wealth of empirical data, many of these studies are not grounded in a clear theoretical framework, nor do their findings consistently contribute to theory refinement. Consequently, while empirical work on tactile localization has multiplied, the relationship between theory and experimentation do not always align and is unclear. The following sections therefore summarize several recent theoretical approaches to tactile localization.

## 2.1 Representational models

Representational models focus on the content of the processed information. They posit that the cognitive system organizes information in (often multiple) map- or container-like units that represent information in specific ways. How information is transformed from one representation to the next is often not specified. Tactile processing, too, has been modeled in this way (Longo et al., 2010; Medina & Coslett, 2016; Tamè et al., 2019). Localization errors in different tasks are then conceptualized as reflecting characteristics of the various specialized representations.

One prominent such model assumes several map-like body representations (Longo et al., 2010; Tamè et al., 2019) which are, however, distorted in characteristic ways (see Fig. 2). For instance, participants make distinct localization errors for tactile stimuli on the hand that are expanded in the mediolateral direction but compressed in the proximodistal direction relative to the veridical hand width and length. Here, these error patterns are interpreted to result from distortions of a presumed "hand map". Reasons for such distortions are sought, for example, in the neural organization of cortical maps, for example in S1, or the shapes of receptive fields. However, distortions in neural maps are larger than those implied by behavioral error patterns; therefore, the model assumes that later representations partly compensate for, or "recalibrate", information from earlier processing stages. For instance, the above model assigns a map (i.e., one container in the model) to the 2D-sheet of the skin, termed the superficial schema. For reaching and pointing to tactile locations, the model assumes two further map-like representations: one that tracks the current body posture, termed the postural schema and yet another one representing 3D body shape, termed the model of body size and shape.

A model like the above-described conceptual model of tactile processing of Longo and colleagues (2010; 2019) can guide experimentation about tactile localization along its propositions. For instance, the postural schema has been experimentally isolated by asking participants to point to body landmarks such as the occluded knuckles and fingertips. Given that no touch is applied with this task, error patterns obtained with this method are then assumed to be touch-independent but posture-dependent and are, thus, ascribed to the respective representation of the model. Other studies have then designed tasks that isolate, according to the propositions made in the model, specific maps, and investigate the assumed progression of information from one representation to the other. For instance, one study (Longo et al., 2015) assessed both localization errors when pointing to body landmarks (posture-dependent) and errors when indicating tactile location on a hand image (body shape-dependent). However, these errors did not add up to the errors obtained when pointing to tactile stimuli on the hand (dependent on both). The model's representations, illustrated as boxes with selective connections between some of them, were then modified because the experimental findings conflicted with the propositions made by the previous model version (Tamè et al., 2019) (see Fig. 2). This strategy of model revision, however, illustrates a commonly acknowledged difficulty of conceptualizing models with boxes and arrows: new findings can typically be implemented by proposing additional representations (i.e., boxes in box-and-arrow diagrams) and connections (i.e., arrows), which limits the models' falsifiability and explanatory ability, so that some rather view them as frameworks than models (cf Press et al., 2022; Pfister, 2019).

Moreover, whether a representation in a model should truly be considered a distinct entity can be a tricky question. For instance, two studies reported that distortions as those seen during pointing to the hand also occur when pointing to other objects based on visual memory (Saulton et al., 2016), and another study attributed distortions to the temporal sequence of responses, with distances being overestimated between consecutive localization judgements that were spatially close (Medina & Duckett, 2017). These findings, thus, cast doubt on whether a distorted hand map truly exists or whether distortions are due to processes that are not specifically related to the representation in question.

Finally, many studies have drawn on the much coarser distinction between two body-related representations, namely a body image and a body schema. These terms are often used with different notions, and discussions persist about the exact characteristics that define the respective representations or additional sub-representations that are posited to collectively make up body image and schema. This situation further highlights how difficult it is to pinpoint exact ideas verbally or as diagrams. It has been criticized that there is a 'chaotic state of affairs' (Berlucchi & Aglioti, 2009), with now many proposals for vaguely defined body representations that are accordingly difficult to test and falsify.

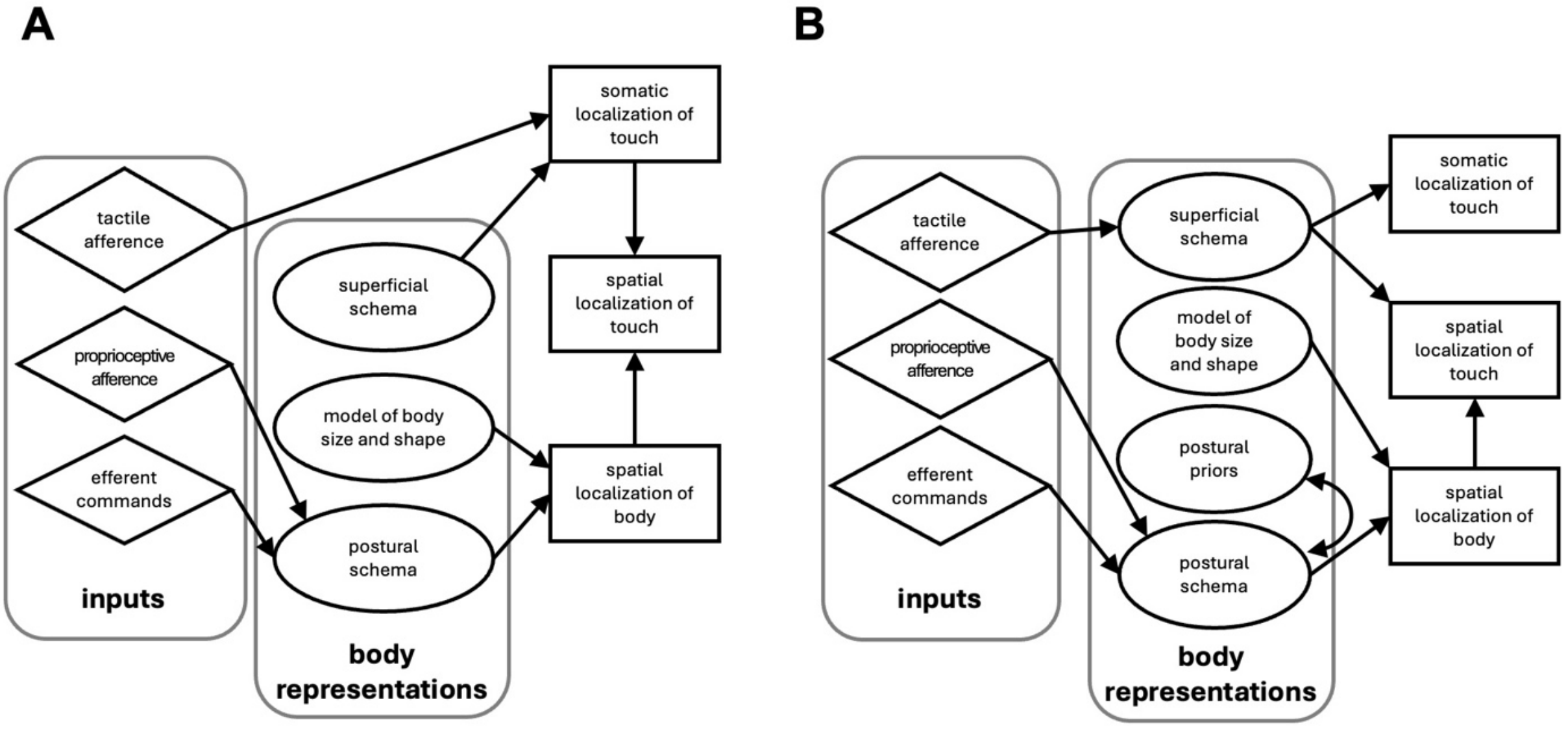

Fig. 2: Representational models of tactile localization. (A) Subset of model by Longo et al. (2010). (B) Updated version of the same model by Tamè et al. (Tamè et al., 2019).

## 2.2 Process-oriented models

As the representational models discussed in the previous section focus on information content, rather than the process of transformation between these representations. Analogously, in box-and-arrow illustrations for these models the arrows between boxes are not defined further than implying the use of information contained in one representation by another. It is a prominent assumption in neuroscience that the brain performs matrix-like transformations between different spatial representations, often designated as “reference frames” (e.g., Deneve et al., 2001; Pouget et al., 2002; Andersen & Cui, 2009). Much work has gone into computational models that is inspired by single neuron recordings in monkey showing that neural processing (i.e., firing of connected neurons) can faithfully implement such transformations (Beck et al., 2008; Ma, 2012; Ma et al., 2006), in line with the idea that errors would be attributable to the resulting map-like representations, rather than the transformation.

Process-oriented models, in contrast, focus on this transformation process more than, or at least equally as, on the underlying representational content. Such models typically offer computational solutions for transformations and, importantly, attribute errors to the transformative process rather than on the representational format.

As an example, recent studies have offered an alternative explanation for the distortions in the presumed maps in representational models. These studies, too, have assumed that pointing responses must rely on localizing stimuli on a limb and then integrating body posture.

However, they posit that distortions arise from optimal integration of sensory information about the body's current and typical, or default, postures (Miller et al., 2022, 2023; Peviani et al., 2024). Default posture is implemented as the prior in Bayesian integration, which has been a popular approach in many areas of neuroscience (e.g., Knill & Pouget, 2004; Goldreich & Tong, 2013; Körding et al., 2007). The critical finding regarding tactile localization is that localization error patterns are explained by a mathematically optimal integration process of undistorted information (Peviani et al., 2024).

On the one hand, this interpretation is based on a similar assumption of that made in the representational models: namely, that different representations, or maps, exist and information is transformed between them. On the other hand, it demonstrates that very different conclusions can be drawn when experimental results are attributed to representations (i.e., the maps proposed by the representational models) vs. to the process of transformation (e.g., a specific way of information transformation or integration). In the representational model view, pointing errors are interpreted as indicating distorted maps. In the Bayesian integration model, the same errors are interpreted to indicate an optimal integration process of undistorted representations of current and prior posture.

An integration view has also been proposed for another set of maps of representational models to account for skin-based localization. Here, it was suggested that touch location is estimated based on the remembered distance of the sensed tactile location on the skin to body landmarks, such as the joints. Just like multilateration works for the localization of a cell phone, based on the phone's distance to several phone masts, touch location may be derived based on distances to body landmarks. Modelling suggested that multilateration explains the curious finding that localization variability is higher for regions of a limb distant from than for regions near the joints (Cholewiak & Collins, 2003; Miller et al., 2022, 2023). As before, rather than interpreting behavioral variability as a sign of map (i.e., representational) deficiency, this theory assigns errors to a statistically optimal integration/estimation process. Again, note that the representational model of the previous section contained a representation that is presumed to contain information relevant to transforming a touch into a location on a body part ("model of body size and shape"), but additional assumptions would be required to account for the modulation of localization variability depending on distance from landmarks. The above process model, in contrast, predicts this feature based on its focus on the process of how information is combined, that is, an arrow in the representational model.

Given these examples, it is obvious that process models may not require some of the representations/maps proposed by representational models. However, one could also imagine casting representational models into a Bayesian framework, effectively specifying connections (arrows) in addition to representations (boxes). Indeed, computational work has explored neural implementations of such information integration (Ma, 2012; e.g., Ma & Jazayeri, 2014).

## 2.3 Tactile remapping and spatial reference frames

Like the process models, the tactile remapping framework encompasses only a circumscribed aspect of the processes considered by representational models, described in the previous paragraph. The tactile remapping framework proposes that all touch is automatically recoded from its original skin-based location into a 3D-external spatial representation by integrating information about skin location and body posture. Thus, different maps and a transformation process between them are again assumed at least implicitly. We outlined earlier that – even if typically presumed – deriving a spatial location of touch is not strictly required theoretically. Hence, the proposal made by the spatial remapping framework that transformation of all touch into a 3D-spatial code is mandatory is neither self-evident nor trivial.

The tactile remapping framework originated in multisensory research (e.g., Shore et al., 2002; Spence et al., 2004): remapping provides a shared spatial code across modalities and so allows aligning tactile input with visual and auditory information. Evidence came from studies showing that the spatial processing of tactile stimuli is influenced by visual stimuli, even when the visual information is task-irrelevant, and, furthermore, this interference is stronger when visual stimuli are spatially close-by than when they are further away from the tactile stimuli (e.g., Spence et al., 2004). The dependence of crossmodal interactions on spatial alignment was interpreted as evidence for touch being remapped into a common space with visual information.

Studies in this research area have primarily implemented forced-choice paradigms (e.g., Yamamoto & Kitazawa, 2001; Shore et al., 2002; Heed & Azañón, 2014), but tactile localization has also been employed. Localization studies have assumed that touch is remapped into a 3D-spatial reference that is eye-centered and lent support from observing that hand pointing towards tactile locations can be biased in the direction of fixation (Harrar & Harris, 2009, 2010; Mueller & Fiehler, 2014a, 2014b). However, eye-centered coding was also reported for proprioceptive (as opposed to tactile) targets (Jones et al., 2010; Mueller & Fiehler, 2014a), so that its use may be triggered by the need to integrate posture rather than being genuinely involved in tactile processing.

Theories of tactile remapping have been extended beyond just an anatomical and a 3D-spatial reference frame. Single cell activity in non-human primate parietal and frontal cortex can reflect various reference frames related to the eyes, head, hand, and torso (e.g., Andersen & Cui, 2009; Batista et al., 1999; Chen et al., 2013). Similar integration of multiple spatial codes has been proposed for tactile remapping, too (Heed et al., 2015, 2016). Moreover, integration may weigh different inputs according to task demands (Badde et al., 2015, 2016; Schubert et al., 2017). Finally, more recent studies have revealed that the body's default posture is probably considered during tactile processing (Badde et al., 2019; Maij et al., 2020;

Romano et al., 2019; Tamè et al., 2019). Accordingly, spatial remapping theories encompass both representational and processing aspects.

## 2.4 Sensorimotor contingency theory

Whereas the three theoretical approaches presented so far all assume some form of representation(s), action-oriented and embodied theories of cognition like the Sensorimotor Contingency Theory (O'Regan & Noë, 2001) reject the idea that representations exist. A key tenet of such theories is the critique, that activity in a certain location on the neural map, like a neuron firing when a stimulus enters its receptive field, does not explain how one localizes the touch (O'Regan, 2010; Wilson & Bednar, 2015). In this view, a relationship between stimulus and neural activity that carries what is typically termed "meaning" can only be established through action. Thus, according to this view, humans establish the lawful relationships of touch perception by movements, for example, when reaching for tactile stimuli.

Such learning does not preclude that representations, or maps, are established by the brain to "represent" the learned relationships. More generally, the term "representation" is not clearly defined in psychology and neuroscience (e.g., Favela & Machery, 2023; Baker et al., 2022). Representational theories, and, in our perception, also the process and remapping theories we have discussed, assume either explicitly or implicitly that some cognitive structures exist that are map-like and can be read out by other cognitive processes. Sensorimotor contingency theories, in contrast, place emphasis on the pivotal role of the interaction of body and environment. In this view, one should not expect, and look for, representations or maps that are independent of action. On a neural level, thus, one would not even expect purely "perceptual" coding even in primary sensory regions of the brain (e.g., Engel et al., 2013).

In our perception, the relevance of action already receives attention in localization research, even if it is typically not explicitly linked to sensorimotor contingencies. For instance, the idea of priors, as formulated in representational, process, and remapping models, easily links to sensorimotor contingencies as the basis of stored priors. Thus, for example, one could frame the Bayesian integration model that explained hand map distortions as a result of integrating a prior as being based on the experienced contingencies of motor responses performed in response to touch, with the prior reflecting the most probable combinations of specific postural configurations when touch is perceived. Following this approach, goal-directed movements, such as reaching towards a touch in external space, need not rely on an explicit representation of that space. Instead, they may be achieved by re-instating previously learned postures that reinstate a set of sensory inputs associated with the desired goal configuration.

Another prediction following from this line of thinking is that participants should perform best when the experimental setup resembles natural, everyday contexts, because experience about typical sensory input exists for them. One line of research has picked up this idea and tested tactile localization with closed-loop feedback. Whereas most studies prevent all tactile feedback during the response to isolate processing of the tactile stimulus, this study explicitly allowed participants to haptically search for the touched target location with their reaching finger (Fuchs et al., 2020), presumably creating an experimental situation that resembles the way we acquire sensorimotor contingencies during everyday life better than typical localization studies. We note, however, that there are alternative accounts for the advantage of closed-loop feedback, not least the fact that such a situation simply contains more sensory information than experiments that prevent such input from occurring. It is our impression that this is a rather general difficulty of the sensorimotor contingency approach: while it is often invoked, it is difficult to pinpoint experimentally.

Viewing all theoretical branches – representational, process-oriented, remapping, and contingencies – together, it is evident that they share many theoretical assumptions and explananda. For example, all four approaches acknowledge the relevance of prior knowledge of the body's default posture and assume some form of transformation between spatial codes. Thus, even if each approach features specific aspects and predictions, it is conceivable that their ideas could be merged into a larger, common framework.

# 3 Typical localization tasks

So far, we've outlined theoretical approaches about how tactile localization might proceed. Here, we will be concerned with screening the literature to assess how research has been conducted. We first use this as a complementary approach to structuring theories. In a second step, we then discuss the identified task categories in the light of their (often implicit) assumptions and, where applicable, of the theoretical frameworks (section 3.4.4).

## 3.1 Methods

### 3.1.1 Systematic literature search

We attempted to collect all published articles on tactile localization fulfilling our predefined inclusion and exclusion criteria, summarized in Table 1.

**Table 1: Inclusion and exclusion criteria**

| Criterion | Included | Excluded |
|---|---|---|
| **Language** | articles published in English language | all other languages |
| **Journal** | • peer-reviewed academic articles<br>• conference proceedings | • non-academic article formats<br>• preprints |
| **Sample** | studies testing non-clinical adult human participants as the main group or as a control group | Studies exclusively testing<br>• non-human subjects, e.g., non-human primates, rodents, other animals, plants<br>• clinical or other special populations<br>• specific age groups, such as elderly, infants, children, or adolescents |
| **Data** | studies presenting primary experimental data | • simulation and modeling without collected empirical data<br>• meta-analyses<br>• reanalysis of elsewhere published data<br>• review articles<br>• commentaries<br>• theoretical articles |
| **Touch** | studies using a tactile localization paradigm, following this general scheme:<br>1. a unisensory touch stimulus is presented on the skin<br>2. participants report in one way or another the location of the stimulus on the skin | • stimuli not tactile, including other somatosensory stimuli like temperature and pain<br>• touch not unisensory, e.g., when stimuli were only presented together with another (e.g., visual) stimulus<br>• touch not presented at a single location, e.g., when tactile distances are presented<br>• touch stimuli presented rapidly, e.g., to invoke spatiotemporal integration or perceptual illusions |
| **Localization Behavior** | studies including a behavioral assessment of tactile localization | • neurophysiological recordings without behavioral localization task<br>• assessment of stimulus location indirectly via another task, e.g., by assessment of reaction times or decisions, such as in the crossmodal congruency task or temporal order judgment task etc.<br>• task is not localization, e.g. judgments of intensity, size, stimulus length, orientation, tactile distance estimation, etc. |
| **Baseline** | studies including a baseline condition where participants' main task was tactile localization, without the influence of experimental manipulations | • effect of drugs, anesthetics or other substances<br>• induction of perceptual illusions<br>• secondary tasks, such as distracting attention or inducing working memory load<br>• any other altered cognitive or physical state, e.g., pain, sleep deprivation, hypnosis, etc. |

We conducted the literature search according to the Preferred Reporting Items for Systematic Reviews and Meta-Analysis (PRISMA) guidelines (Moher et al., 2009; Page et al., 2021). The guidelines propose the use of a checklist of recommended items to be reported and a description of the selection process using a PRISMA flow diagram with four steps (see Fig. 2). We describe the four steps of our selection process in the following.

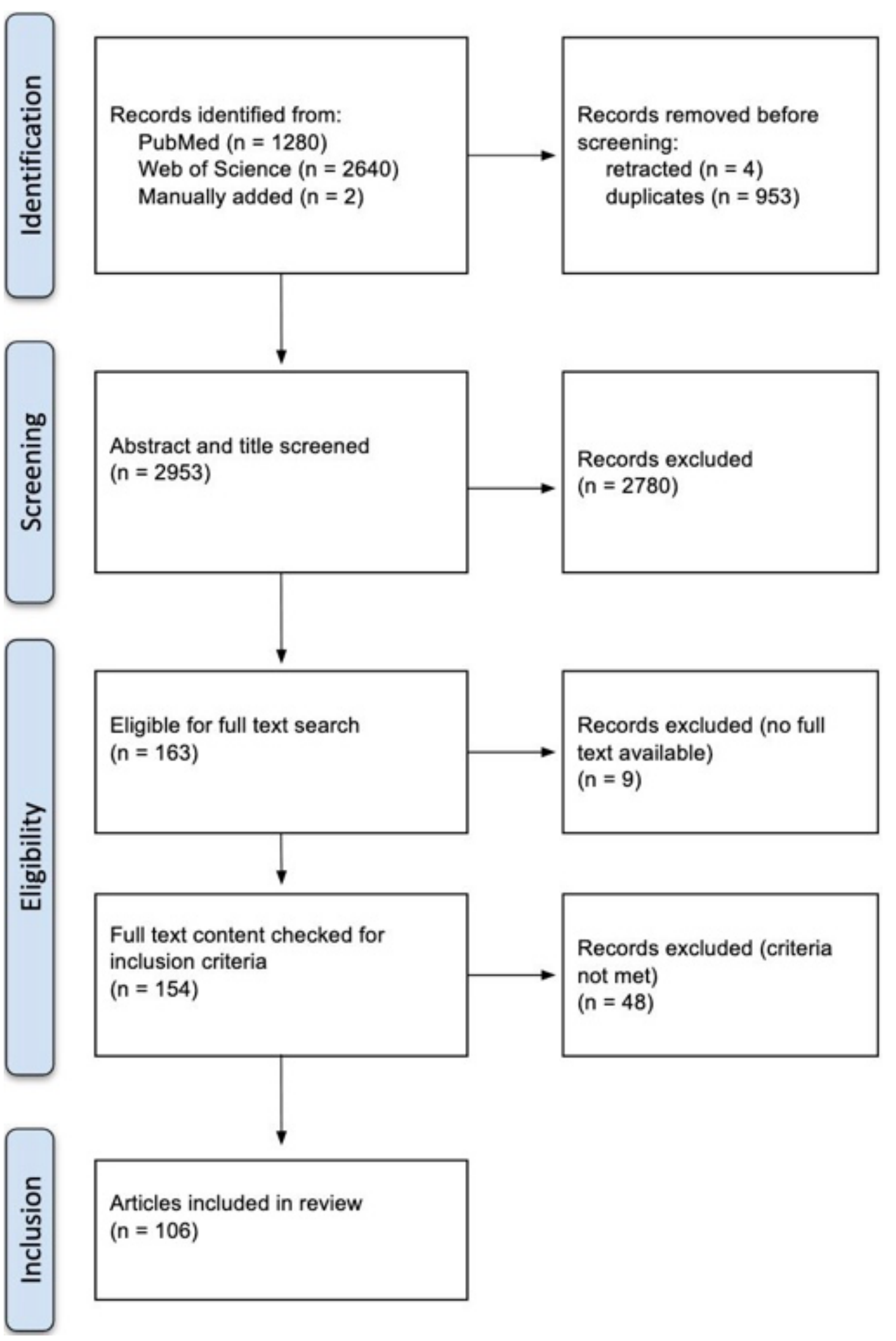

Fig. 3: PRISMA flowchart for describing the literature search and selection process.

### Step 1: Identification

We searched within the databases PubMed (https://pubmed.ncbi.nlm.nih.gov) and Web of Science (https://www.webofscience.com) and included all search results until the day of the search, which was September 7th, 2023. For PubMed, we used the search term *("localization" OR "localisation") AND (tactile OR touch*).* For Web of Science, we used the search term *(localization OR localisation) AND (tactile OR touch*)*. We combined the records obtained from PubMed and Web of Science and removed duplicates and publications marked as retracted. After comparing the search results with the literature known to the team, we added two studies manually. One study referred to localization as “position sense” (Longo & Morcom, 2016) and the other mentioned “mislocalization” and “positions on the body surface” (Trojan et al., 2010).

### 3.1.2 Step 2: Screening

Screening and extraction were conducted by three of the authors. XF screened studies with the first author’s name starting with letters A to G, JK with letters H to P, and SP with letters Q to Z. Any cases that were unclear were discussed between these authors and classified through consensus. In the initial screening phase, articles were evaluated based on the information provided in the title and abstract. Any articles that clearly did not meet the inclusion criteria were discarded. Ambiguous items were then evaluated in the subsequent stage.

### 3.1.3 Step 3: Eligibility

For the remaining items, we attempted to obtain the full text of all identified papers. We were able to obtain 154 of 163 full texts.

### 3.1.4 Step 4: Inclusion

We screened the full texts and selected those that matched the seven inclusion criteria (see Table 1). 106 papers matched all 7 criteria and entered the data analysis. Some papers included more than one tactile localization experiment (see the results section for details) and in this case we extracted the data for each experiment separately.

## 3.2 Information extraction and study characterization

We extracted the information listed in Table 2 from each paper for further analysis.

**Table 2: Extracted information from the selected experiments**

| # | Category | Details and subcategories |
|---|---|---|
| 1 | Author | Name of first author |
| 2 | Year | Year of publication |
| 3 | Experiment number in article | If several experiments in a study fulfilled the inclusion criteria, this variable indexes these experiments. |
| 4 | Sample | Sample size and type, e.g., “20 healthy adults” |
| 5 | Description of experiment | • Type of experiment (categorical, e.g., “psychophysiological study”)<br>• Brief description (free text description, e.g., “effect of hand crossing on touch localization of finger”) |

| 6 | Type of touch stimulus | • Physical quality of stimulus (categorical, e.g., “vibration”)<br>• Type of device (categorical, e.g., “monofilament”) |
|---|---|---|
| 7 | Location of touch stimulus | • Stimulated body part (free text description, e.g., “palmar surface of left hand”)<br>• Stimulated body area (categorical, e.g., “hand”)<br>• Number of stimulated locations and stimulus arrangement (free text description, e.g., “7-stimulator linear array”) |
| 8 | Localization method | • Free task description (e.g., “mouse click on perceived position on digital hand image”)<br>• Spatial response: Participants were instructed to respond spatially, e.g., point at a location (yes/no)<br>• Anatomical, categorical response: the task explicitly requires participant to respond about a body part, like naming a touched finger (yes/no)<br>• Task requirement (categorical, e.g., “indicating a location”, “deciding between options”)<br>• Reporting method (categorical, e.g., “reaching/pointing”)<br>• Spatial stimulus-response relationship: the type(s) of spatial transformation(s) necessary to transform the stimulus space to the response space (categorical, e.g., “translation”)<br>• Reference (categorical, whether responses were given in reference to another stimulus or landmark, i.e., “relative”, “absolute”)<br>• Response alternatives or degrees of freedom in the responses (categorical, e.g., “continuous”, “discrete”)<br>• Dependent variable (categorical, e.g., “accuracy”) |
| 9 | Main result(s) | Brief description of main results (free text description, e.g., “distal bias for stimuli close to the wrist”) |

## 3.3 Statistical analysis

### 3.3.1 Attributes of tactile stimulation

To analyze which stimulation methods are most common by using variables 6 and 7 from Table 2, we determined the physical quality of the stimuli and computed the frequency of their use across studies. For the most common categories, we also computed the frequencies of the use of specific stimulation devices. “Location of touch stimulus” was analyzed using the same approach by coding body segments and computing the frequency of studies stimulating the respective segments. We extracted which segments were stimulated and, for simplicity, combined ventral and dorsal parts of the segments.

### 3.3.2 Attributes of localization tasks

We explored whether localization tasks fall into distinct categories. We selected five key task attributes as classification variables and analyzed the frequency of their combinations. We conducted this analysis at the level of individual tasks rather than experiments, because some experiments included more than one localization task. The five key attributes were derived from the "localization method" section (see variable 8 of Table 2), applying a hierarchical (arborescence) classification scheme: (1) spatial response, (2) spatial stimulus-response relationship (i.e., spatial transformation), (3) task requirement, (4) reference, and (5) reporting method.

# 4 Results

Our literature search yielded 106 eligible articles that were analyzed for the review. The results comprised a total of 171 experiments and a total number of 2949 data sets from healthy participants; in one experiment (Craig & Rhodes, 1992) the number was not reported. Note that some studies tested the same participants multiple times, so the reported number is slightly higher than the number of unique individuals.

Fourteen experiments included two different localization tasks. Some information was missing for the task aspects. For one study (Hsu & Nelson, 1981) the task was not clearly described and it was excluded from the analysis. For three studies (Nguyen et al., 2020 (two tasks); Schlereth et al., 2001; Ylioja et al., 2006), not all variables included in this analyses were available and thus, the resulting arborescence classification presented in Fig. 4 is based on 180 out of 184 tasks. information was not available. Hence, 184 localization tasks coming from 170 experiments were included in our analyses. Note that in some of the following analyses we present proportions that are either relative to the number of experiments or the number of tasks.

All included studies are listed in the Supplementary Material. The extraction results for all studies are shown in Table S1, provided as online Supplementary Material.

### 4.1.1 Attributes of stimulation

With only very few exceptions (e.g., one study not reporting the stimulus and one using a brushing type of stimulus) the physical stimulus quality of all experiments could be classified into the categories of *pressure*, *vibration*, or *electrical* in a straightforward manner. Note, some experiments used more than one type of stimulus; therefore, the reported numbers sometimes exceed the number of experiments. The results are illustrated in Fig. 3A.

The majority, i.e., 102 (60%) of the 171 analyzed experiments used pressure stimuli; 55 (32.4%) used vibration stimuli, and 14 (8.2%) used electrical stimuli. For pressure stimuli, the most used stimulation device was a manually applied von Frey monofilament (48 of 102 experiments using pressure stimuli, 47%) followed by other hand-held objects (18 experiments of 102, 17.6%), followed by use of the experimenter's finger (10 experiments of 102, 9.8%).

Hence, computer-controlled stimulators that were attached to the skin throughout the experiment, such as solenoids, were relatively rarely used to induce pressure stimuli (6 out of 102 experiments, 5.8%).

To induce vibratory stimuli, the variety of applied devices was higher. In the subset of the 55 experiments using vibratory stimuli, the most frequently used devices were electromagnetic tactors (15 out of 55 experiments, 27.3%), rotating mass motors (14 out of 55 experiments, 25.5%), and piezoelectric stimulators (10 out of 55 experiments, 18.2%).

### 4.1.2 Stimulated body areas

The stimulated sites could be attributed to belong to any of the following segments of the body: fingers, hands, lower arm, upper arm, neck, head, face, tongue, chest, upper back, lower back, abdomen (or belly), thighs, lower legs, feet, and toes.

The segments which were most frequently stimulated in tactile localization studies were the upper extremities (see Fig. 4B). The highest frequencies were found for the forearm (58 out of 171 experiments, 33.9%) as well as the hand and the fingers (each 46 experiments, 26.9%) followed by the abdomen with, however, already considerably fewer occurrences than for the former (10 experiments, 5.8%). For most other areas of the body there were very few studies (see Fig. 4B) or even none (e.g., for the thighs).

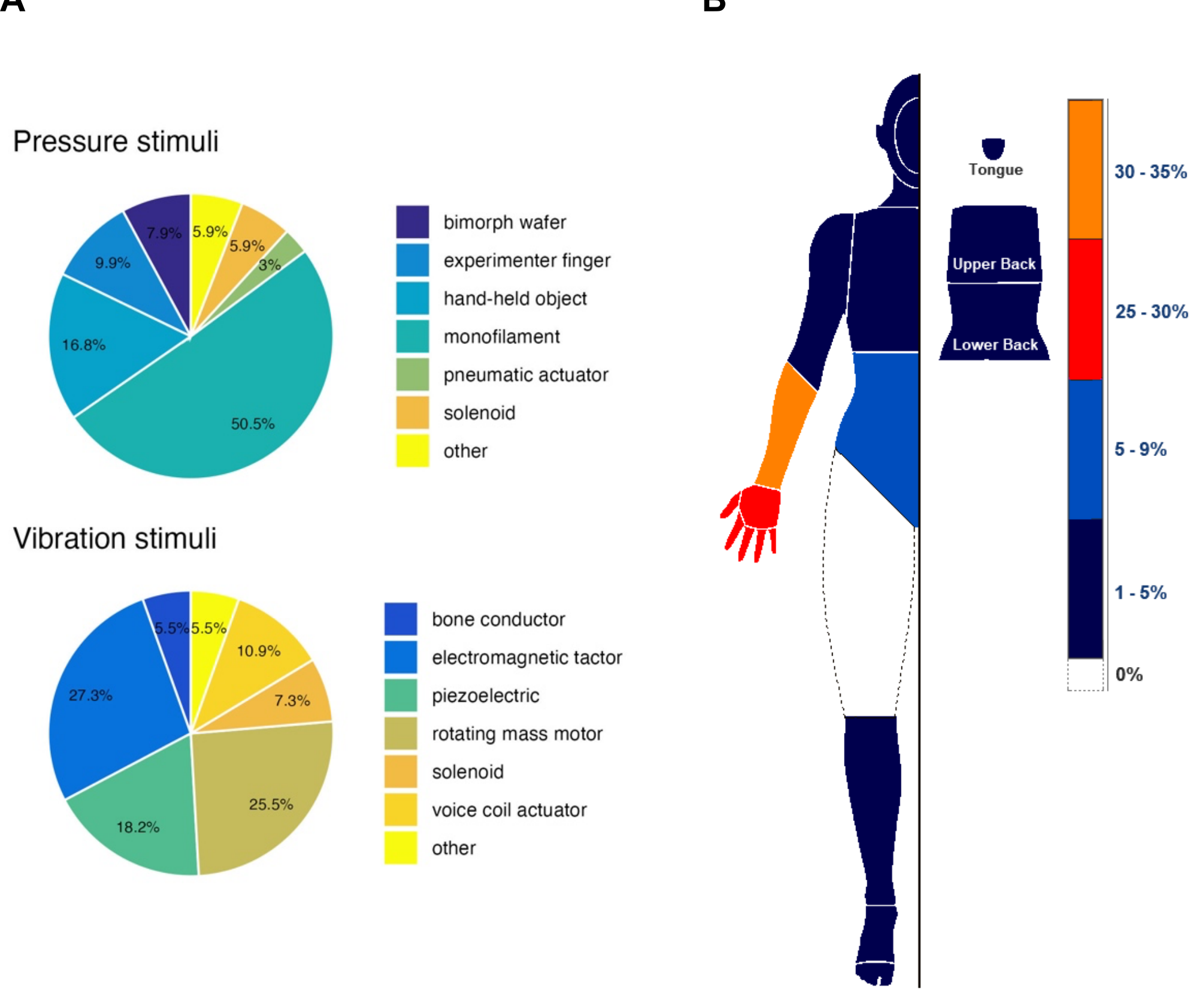

Fig. 4: A: Proportions of stimulation devices used for applying pressure and vibration stimuli in localization studies. B: Proportions of experiments testing various segments of the body.

### 4.1.3 Attributes of the localization task

We identified a set of task categories that reflect general methodological approaches to studying tactile localization based on whether the responses were spatial, whether spatial transformations were required between stimulus and response, the type of task performed, the presence of spatial references, and the type of reporting method used. Before discussing this classification in detail, we will first describe each of the identified task features individually.

#### Task requirements

The most obvious split in task characteristics was that participants either *took decisions* (93 out of the 180 tasks, 51.6%) or *indicated* a touched location (85 out of 180 tasks, 47.2%). Only two tasks (1.1%) could not be assigned into one of these two categories and are described as "other" (Benedetti, 1988; van Erp, 2008).

#### Reporting method

Across all task categories, we discovered these reporting methods: reaching/pointing (56 of 180 tasks, 31.1%), verbal responses (45 tasks, 25%), button presses (36 tasks, 20%), show location on an image (34 tasks, 18.9%), move the touched body part (4 tasks, 2.2%). For only 5 tasks (2.8%) the reporting method did not fit into these categories (Benedetti, 1988; Eguchi et al., 2022; Mancini et al., 2011b; Sadibolova et al., 2018; van Erp, 2008).

#### Spatial vs. non-spatial response requirements

Reaching and pointing responses are, of course, always inherently spatial. However, when participants make decisions about tactile stimuli, responses can be spatial or non-spatial. For instance, if participants indicate their response by pressing one of several buttons, this response is again inherently spatial, because the button press occurs in space. In contrast, responses can be devoid of any spatial aspects, for example when participants name the finger that has been touched. Most response requirements were spatial (155 out of the 180 tasks, 86.1%) and only few were non-spatial (25 tasks, 13.9%).

#### Absolute vs. relative response requirements in decision tasks

When participants had to make decisions (93 out of 180 tasks, 51.7%), these could be either *absolute* (81 out of the 93 decision tasks, 87.1%) or *relative* (12 tasks, 12.9%). For absolute decisions, participants reported one of several stimulus positions without any explicit reference point. For relative decisions, participants reported on a stimulus position in reference to another stimulus or an anatomical landmark.

#### Spatial relationship between stimuli and responses

Whenever participants are required to give a spatial response (155 out of the 180 tasks, 86.1%), they must map the stimulus position into the spatial code of the response. This implies the potential for (in)compatibility between the two spatial locations. For instance, a button on the left would be spatially (in)compatible with a stimulus on the left (right) side of the body. Such (in)compatibilities can be more or less obvious. We identified, for each experiment, the number of spatial transformations, such as translation, rotation, scaling, and inversion, that are technically required to relate stimulus and response. This identified three broad categories: tasks that did not require any spatial transformation (47 out of the 155 spatial tasks 30.3%) because the response was in the same space as the stimulus; tasks with only translation, either vertically or horizontally (47 tasks 30.3%); and tasks requiring multiple transformations (61 tasks, 39.3%). Examples of such tasks can be seen in Fig. 5.

### 4.1.4 Categories of localization tasks

Our main aim for analyzing the attributes we have discussed so far was to cluster tasks into categories (see Fig. 5 and 6) by combining the aspects described above. Indeed, we identified 10 such categories that we will discuss in detail below. Each method may potentially recruit different sensory information and specific cognitive processes during the response. This emphasizes that what is commonly referred to as “tactile localization” in psychology and neuroscience is not one clearly defined, isolated sensory process but rather a heterogenous family of various testing contexts. We would like to highlight that although the 10 categories (out of 31 branches of the arborescence tree) describe important variants of tactile localization tasks, the 10 categories accounted for 99 of the 180 tasks (55%) which again underlines the large heterogeneity of tactile localization methods.

Studies that require a spatial response fall roughly evenly into three categories: they require no spatial transformation, they involve a simple translation, or they necessitate multiple spatial transformations of the tactile location to make the response. Tasks that do not require a transformation of tactile location, or that only require a translation (e.g., pointing to a location above the touched arm) typically require a reach-to-point response. Accordingly, these tasks typically implement absolute localization demands and not localization relative to another landmark.

Moreover, some response strategies may involve localizing touch on a body part independent of its posture and relation to the remaining body, whereas some response strategies may, instead, use 3D-like, spatial coding or postural information. For instance, when a touch on the forearm must be localized on a drawing of the forearm, it is sufficient to determine the distance of the touch between the elbow and wrist and then indicate an analogous location on the drawing. It is not necessary to derive a spatial location of the touch in 3D space and somehow transform this location to match the drawing. We will refer to such localization strategies as body part-centered or, in the case of the limbs, as limb-centered; conceptually, they may be considered as relating closely to anatomy, because no 3D-spatial localization must be performed to solve the task. Such localization strategies contrast with those that involve postural information. For instance, when the left arm is hidden under a board and the right arm must point on the board above the touched location, then simply deriving touch location relative to elbow and wrist is not sufficient; instead, posture must be integrated to derive where the right hand should be directed. In cases like these, we will refer to spatial localization strategies.

By contrast, tasks that involve multiple spatial transformations, such as combinations of translation, scaling, or inversion, more often rely on categorical decisions rather than direct spatial actions. In such cases, participants usually report the location via button presses rather than goal-directed movements toward the stimulated body part.

The relatively small subset of studies that do not involve spatial responses at all typically employ verbal reports and focus on absolute decisions, such as naming the touched location without requiring any spatial transformations.

Given the large variability in localization tasks, it was not always possible to precisely categorize each study. For instance, we categorized one study in which participants pointed onto their arm with a laser pointer (Samad & Shams, 2016, 2018) as “direct pointing to an image” without spatial transformation. To allow readers to retrace our categories, the references to all studies falling into each branch of Fig. 4 are listed in the Supplementary Material. Task categories are illustrated in Fig. 4, which presents an arborescence classification based on our five task attributes. In the following, we will describe each category in more detail.

Fig. 6 depicts the most common categories employed for tactile localization experiments, as identified in the previous section (see Fig. 5). We will briefly discuss each of these methods in the following sections.

All studies reporting tasks falling in the categories of the arborescence classification tree are listed in the Supplementary Material (see section 2 of the Supplementary Material and Table S2).

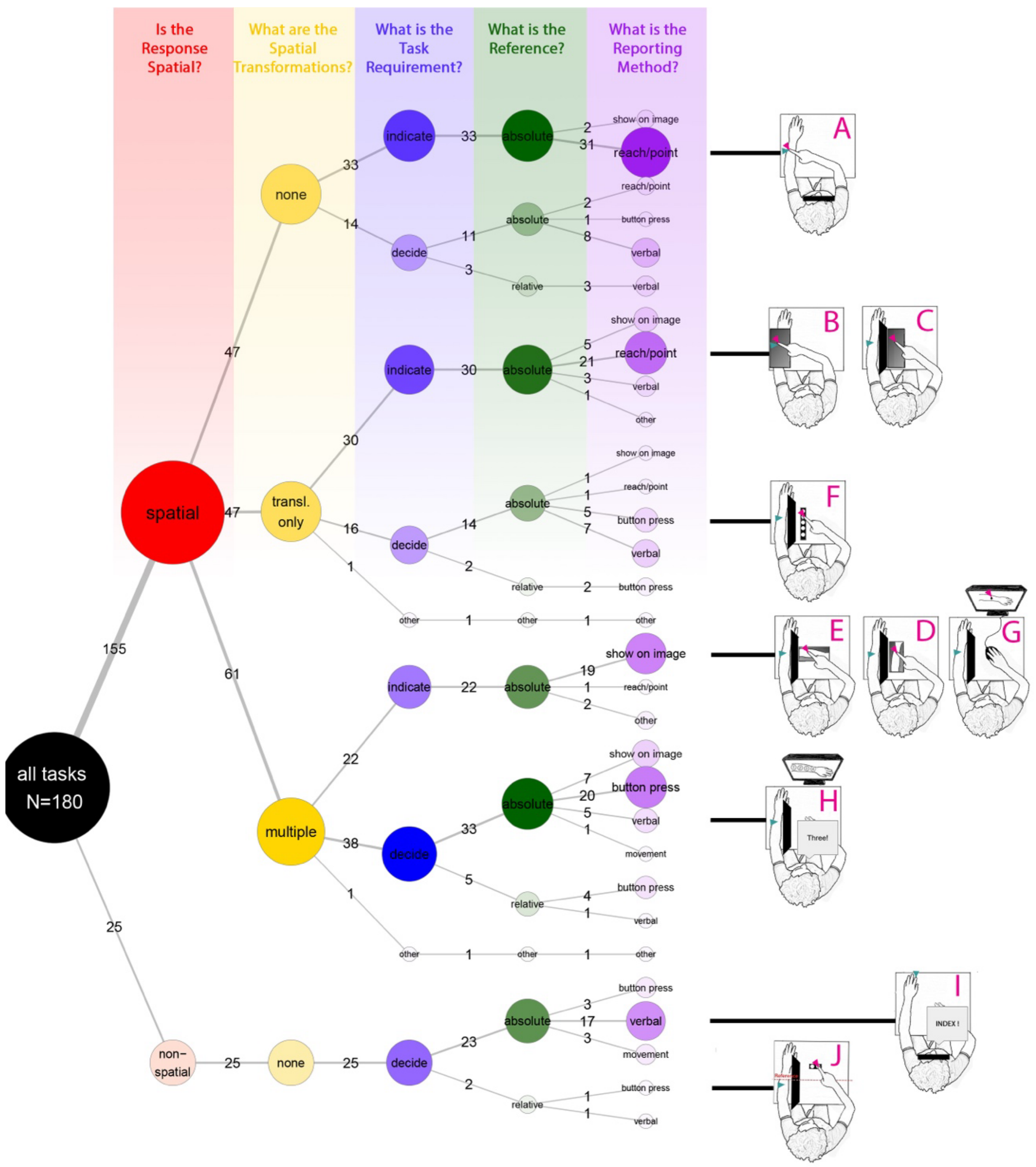

Fig. 5: Arborescence classification of localization task aspects with the layers (1) spatial response, (2) spatial stimulus-response relationship (i.e., spatial transformation), (3) task requirement, (4) reference, and (5) reporting method The numbers of tasks for different combinations of aspects are indicated by the sizes and luminance of the circles and the exact numbers are shown along the edges of the tree. As can be seen, the numbers differed considerably between the combinations. Some combinations with relatively high frequencies create clusters representing typical localization tasks. We highlighted those in the rightmost column part of the figure showing thumbnails of typical implementations of the types of task that we will describe in detail in the following sections.

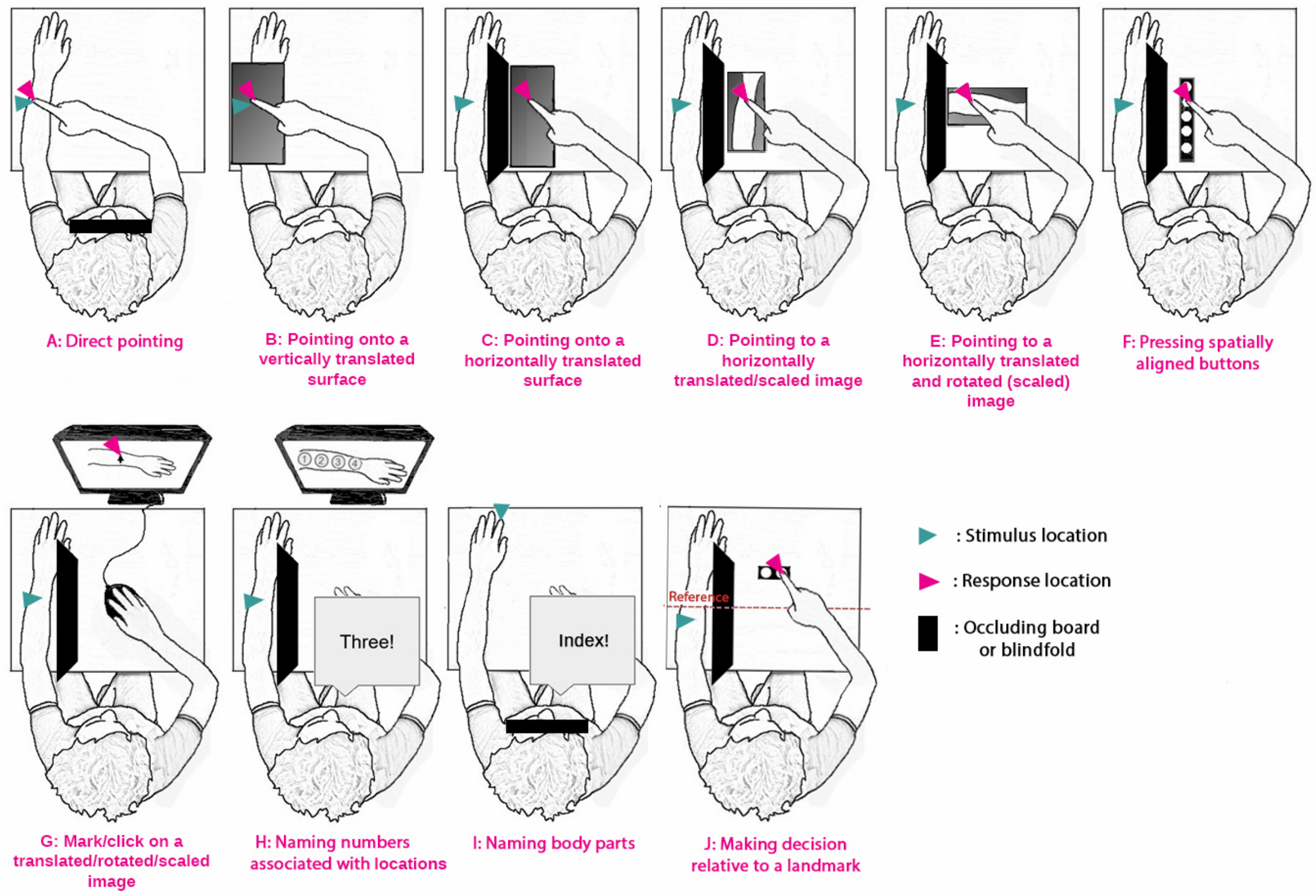

Fig. 6: Illustration of the most common methods of tactile localization.

## Direct pointing

Direct pointing tasks (Fig. 6A) require participants to reach or point directly to the stimulus location on the skin (Fuchs et al., 2020; Medina & Rapp, 2014). These closely resemble natural behaviors, such as swatting a mosquito, as they present stimulus and response in the same space. While their ecological validity is an advantage, their interpretation is complex due to the variety of strategies participants might employ.

Direct pointing tasks appear to engage remapped, external-spatial representations, given their reliance on a spatially directed movement and the absence of an explicit anatomical reference. Yet, they do not always preclude body part-based strategies. Instead of fully relying on an external spatial code, participants may code the location relative to an anatomical landmark and infer its spatial coordinates in a second step when necessary. This strategy is particularly plausible when visual information about the limb is available (Fuchs et al., 2020). Another possibility unique to this mode of localization is that participants may use tactile-tactile feedback and compare the remembered tactile stimulus to newly generated input from touching their own skin with the responding finger. In this case, precise spatial remapping or a clear representation of the body may be unnecessary; a coarse spatial estimate may suffice, with fine localization achieved through closed-loop feedback, essentially using the body “as its own best model" as the sensorimotor contingency framework would suggest (cf Fuchs et al., 2020).

## Pointing onto a vertically displaced surface

Many studies prevent skin contact during localization to eliminate closed-loop feedback associated with direct pointing. The rationale for this experimental strategy is that localization is considered the process that derives a spatial location purely from the tactile stimulus alone, without any further, self-produced tactile input. Participants either point without touching (e.g., Brandes & Heed, 2015), they point to a location directly above the perceived location (e.g., Trojan et al., 2010), or they point onto an occluding surface placed directly above the target region with their index finger or a hand-held tool, like a baton (e.g., Longo et al., 2015; Fuchs et al., 2020; Hidaka et al., 2020a; Medina & Duckett, 2017). In these cases, stimulus and response spaces are not spatially identical, but only a vertical translation is required between them (see Fig. 6B).

Although this displacement of the response relative to the stimulus may appear minimal, its effects on localization are not well understood. It seems to us that it is often implicitly assumed that responses are merely shifted uniformly along the vertical axis. However, this may be simplified. If participants aim toward the actual stimulus location, instead of a location on the barrier that is perpendicularly above the stimulus location, localizations can be biased. One study, for example, showed that when pointing movements approached from the side, responses were shifted in the direction of the movement's starting point (Fuchs et al., 2020). This may indicate that participants had not aimed at the location above the hand stimulus on the occluding board, but that a reach directed to the true stimulus location on the hand had been intercepted in flight by the occluding board. The extent to which participants may adapt to such constraints over time is unknown.

Moreover, this method shares interpretive ambiguities with direct pointing. Although the task is typically framed as spatial, anatomical strategies, such as aiming for a proprioceptive finger location rather than the stimulus proper, remain a potential alternative.

## Pointing on a laterally displaced scale

Closed-loop feedback can also be prevented by having participants point to a laterally (rather than vertically) displaced surface (e.g., a scale, tablet or rail positioned beside the arm; see Fig. 6C) (Brooks et al., 2019; Miller et al., 2022), while no image of the respective body part (e.g. the hand or arm) is shown on the response surface. This setup is typically one-dimensional, that is, participants indicate, say, the proximal-distal location of a stimulus along the length of the arm. Although the task still involves a spatial translation, the physical separation between stimulus and response spaces makes it impossible that participants directly reach toward the stimulus and are interrupted by a barrier. Instead, they must intentionally direct their response to a space that is distinct from that of the stimulus region.

In this context, participants could either translate their intended pointing direction to match the displaced surface for each individual pointing response; instead, they may attempt to map the stimulus space onto the response space as a whole. Such considerations are important when one wants to interpret errors, as they may be related to motor planning (for individual pointing responses) or incorrect translation between stimulus and response spaces. Moreover, different theories would likely make different assumptions about how participants solve this task. The remapping framework would suggest that participants determine the 3D location of the tactile stimulus on the arm and translate this location onto the scale. In contrast, picking up the notion of landmarks proposed by the multilateration model, participants might identify the distance of the tactile stimulus from one or several landmarks, localize the landmark(s) on the scale, and then derive stimulus location on the scale. It is not clear whether such different cognitive implementations would lead to distinct error patterns; our point is, however, to demonstrate that interpreting a seemingly simple localization task is not straight-forward.

## Pointing to a (scaled) image aligned with the arm

The lateral displacement can be complemented by a graphical representation of the body (part), such as a drawing, photograph, or silhouette (Medina & Duckett, 2017; Miller et al., 2022). It is not clear how participants match the tactually perceived location on their arm to the visual representation. Thus, the relationship between stimulus and response is not transparent, making the underlying localization processes difficult to interpret. The visual alignment of the representation may encourage limb-centered localization by providing an interface that visually displays and emphasizes anatomical landmarks. When the representation is matched in size to the participant's own limb, localization can in principle rely on either limb-based (e.g., landmark-based) strategies, or on spatial strategies, with only a simple translation required to map between the stimulus and response spaces.

One study (Miller et al., 2022) has, however, purposefully rescaled the image in the localization task to enforce that participants must "represent" the touched area rather than attempt spatial matching. While a rationale is given for the use of rescaling, it is unknown how exactly this rescaling affects the localization process. On the contrary, another study attempted to avoid incongruencies between the presented visual image and the (external spatial) location of the limb and showed an arm image suspended above the real arm (Steenbergen et al., 2013, 2014). While there is also a rationale for this choice, this approach leads to ambiguity by allowing various strategies, or even a mixture of strategies, to solve the task.

## Pointing to a (translated/rotated/scaled) image

This method differs from that of the previous section in that the visual representation of the body is shown in a spatially unrelated location, such as on a monitor, and responses are made using an input device such as a mouse (Kang & Longo, 2023; Mancini et al., 2011a; Martel et al., 2022) (see Fig. 6E). Participants are typically instructed to click on the location that corresponds to where they felt the touch. One version of this approach involves dividing the skin into labeled grid sections and asking participants to report the matching area (e.g., Weinstein, 1968).

The use of an image may primarily promote a body part-centered response strategy: A purely spatial strategy would imply translation, rotation, and scaling of the tactually perceived location to derive the response. Instead, it seems likely (though proof is lacking) that participants first localize the touch relative to the touched body part, such as relative to the wrist and elbow for a lower arm stimulus; second, they then identify and report the corresponding location on the image based on characteristics of the image, e.g., in the previous example relative to wrist and elbow location depicted on the image. Notably, having identified the tactile location anatomically, translating it to the misaligned response image likely induces spatial processes nonetheless. This creates the interpretational difficulty that it becomes more difficult to tell whether any spatial influences stem from a body representation or from the need to prepare a spatial response, for example, by mentally rotating a "hand representation" to the image of the hand. Difficulties may arise both for representational accounts (which and how many representations are invoked?) and process-oriented accounts (which process is affected?).

Yet another difficulty arises from the fact that researchers typically do not specify whether they think that their experimental manipulation affects a full-body representation or a representation restricted to the tested body part. Many discussions fall back on terms like "body schema" that invoke a full-body representation. However, many papers propose, for instance, "hand maps" or "hand-centered", "trunk-centered", and "head-centered" reference frames. Notably, the spatial transformation of a hand map into a hand image is different than the spatial transformation of the entire body in such a way that, afterwards, the hand would be aligned with the hand image. Thus, interpretational ambiguities arise from underspecification of the assumed underlying representations targeted by the experiment.

## Pressing spatially aligned buttons

In this task, participants localize stimuli by pressing one of several physical buttons arranged alongside the body part (Cholewiak & Collins, 2003; Cholewiak & McGrath, 2006). The setup shares similarities with pointing to a laterally displaced surface but constrains responses to discrete options rather than a continuous 2D space. There is ample evidence for the existence of stimulus-response compatibility effects (e.g., Schubert et al., 2017; Riggio et al., 1986). Therefore, the button arrangement may encourage a spatial localization strategy that translates the spatial location of the tactile stimulus to button, but we are not aware of any empirical test of this suggestion. Accordingly, as with other spatial tasks, it is possible that participants use limb-centered strategies; this possibility may be more relevant the more complex or difficult the alternative spatial transformation. Moreover, it is conceivable that participants, at least over the time of an experiment, categorize the stimuli ("this is the second of the five stimuli in this experimental setup"), and it is unclear which cognitive strategies would be implied by such strategies.

### Naming body parts

In this approach, participants verbally identify the body region where they felt the stimulus (Braun et al., 2005; Harrar et al., 2013), either by freely uttering the word or by selecting from labeled response options. This task is constrained to locations that can be distinctly named, such as fingers and knuckles, so that the response does not require any spatial mapping. However, even in such tasks, participants may implicitly use spatial information to infer the correct anatomical label.

However, different from all tasks so far, naming body parts requires that specific parts of the skin are matched to a categorical response. It is not clear how our categorical names for body parts are derived, though sensorimotor contingency theories would posit that (de-)correlations would identify body parts with experience. Moreover, perceptually relevant borders between body parts do not necessarily match with those for our language. For instance, localization across the wrist affects how we perceive distances between two tactile stimuli (de Vignemont et al., 2009); this is the case even for languages that do not differentiate between arm and hand (Knight et al., 2020). Last but not least, participants are prone to making categorical errors when assigning touch to a limb, such as confusing hands and feet (Badde et al., 2019; Heed et al., 2024). How all of these aspects should map onto body representations, and how our cognitive system mediates between them, is not known. Thus, a seemingly simple form of response introduces a plethora of potential confounds, and the invoked cognitive processes are all but clear.

### Making decisions relative to a landmark

Here, participants localize touch relative to a reference point, such as reporting whether a stimulus occurred to the left or right of a predefined landmark or reference stimulus (e.g., Cody et al., 2010; Azanon et al., 2010; Pritchett et al., 2012). Responses are typically made via button presses. This task likely favors non-spatial strategies by explicitly tying localization to the structure of the body. However, the spatial arrangement of the response device can play a significant role. For example, if participants press a more distal button for a more distal stimulus, spatial mappings may still be formed between external space and response layout – blurring the boundary between anatomically based or limb-centered and spatial strategies.

# 5 Quantitative comparison of localization methods

Given the multiplicity of studies and experimental protocols reported in the literature, it might be difficult for researchers to compare the findings and get relevant insight from them. To facilitate this process, we proposed visualizing the main findings of several studies in a common frame of reference, making them visually comparable. We selected the studies according to three criteria:

1. Data Availability: The data points (mean responses or individual data points) should be accessible directly from the article, through a data repository, or as figures from which individual data points or summary statistics (mean values and/or variable errors) can be extracted.
2. Stimulus Location: As most studies focus on the forearm and the hand (see Fig. 4), we decided to compare the results according to those locations. Hence, our second criterion was to only include experiments conducted on the forearm (dorsal surface) or the hand (dorsal surface).
3. Response type: While our goal was to compare different methods, the response produced by the participant should be spatial (see Fig. 5).

We identified 10 studies that corresponded to those criteria. We used WebPlotDigitizer (https://apps.automeris.io/wpd4/) and a custom program developed in MATLAB (version R2023b) to extract and project the data on an image of a forearm used in Martel et al. (2022) and an image of a hand. A detailed description is available in the Supplementary Material. We show the bias and variability from three studies for the forearm (Brooks et al., 2019; Fuchs et al., 2020; Martel et al., 2022), and the bias from seven studies for the hand (Hidaka et al., 2020b; Kang & Longo, 2023; Longo et al., 2015; Mancini et al., 2011a; Margolis & Longo, 2015; Medina & Duckett, 2017).

We have argued that each experimental method subsumed under the umbrella term "tactile localization" potentially evokes specific cognitive processes. Each distinct process may invoke its own (additional) error pattern. Vice versa, if different experimental methods resulted in distinct error patterns, this would support our conjecture. Therefore, we visualized a quantitative comparison of the localization errors reported for different methods. Our comparison focused on the hand dorsum (Fig. 7) and dorsal forearm (Fig. 8). We mapped the responses of all scrutinized studies onto the same hand image. We selected studies based on the following criteria (see Supplementary Material for details); (i) stimulation of the hand or forearm, (ii) spatial error was reported as a dependent variable, and (iii) the required data were either available or could be derived from a figure in the respective paper.

### 5.1.1 Extraction for graphical comparison

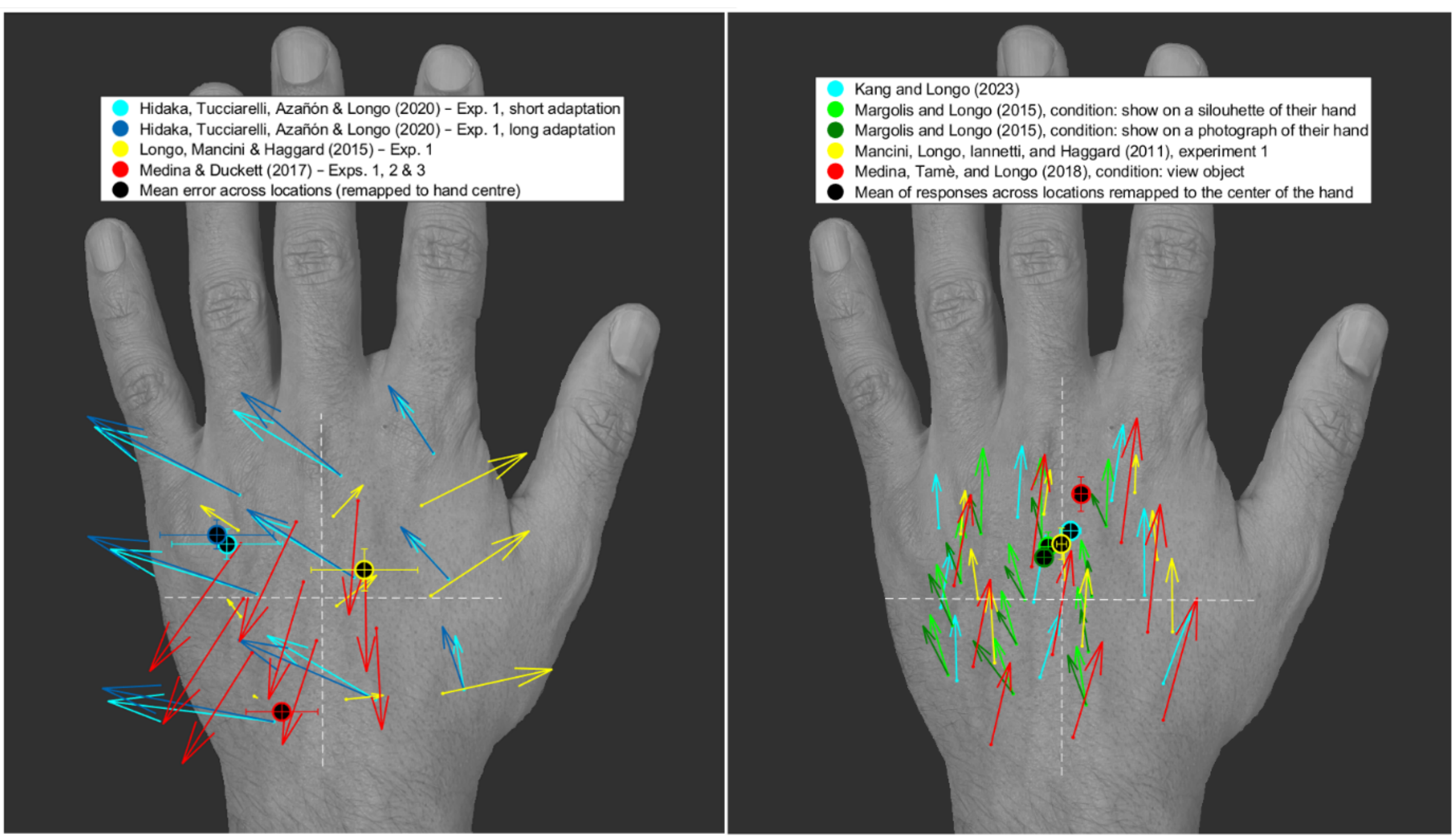

Fig. 7: Bias (constant error) of responses for tactile localization on the hand. On the left, participants responded by indicating perceived stimulus location with a baton on an occluding board placed over the stimulated hand. On the right, the participants responded by clicking on an image or a silhouette of the hand.

Fig. 7 depicts the bias of localization responses on the dorsal hand reported by seven exemplary studies that employed two different localization tasks.

#### Bias

The left panel of Fig. 7 shows results from three different studies (four different experiments). In the original articles, the results were shown after applying transformations to the localized positions. In Hidata et al.  (2020b) and Longo et al. (2015) the Procrustes alignment method was applied to focus on the shape of the localized pattern by removing shifts in location, rotation, or size. In the figured by Medina and Duckett (2017), the responses were recentered over the original targets, thus removing constant error biases (proximal and ulnar – which were reported). All these studies paint a coherent picture in terms of the perceived hand shape – wider and shorter than it really is. However, when preserving the constant errors (left panel of Fig. 7), we observe that there are (i) large constant errors (biases), and (ii) they are largely inconsistent between the different studies.

The right panel of Fig. 7 depicts pointing error when participants had to indicate stimulus location on a hand image with the computer mouse. All responses exhibit a distal bias, that is, responses are biased towards the fingers, even if the size and the specific direction of this bias varied among studies.

## Variability

Due to the absence of available data, the variability is not reported in Fig. 7.

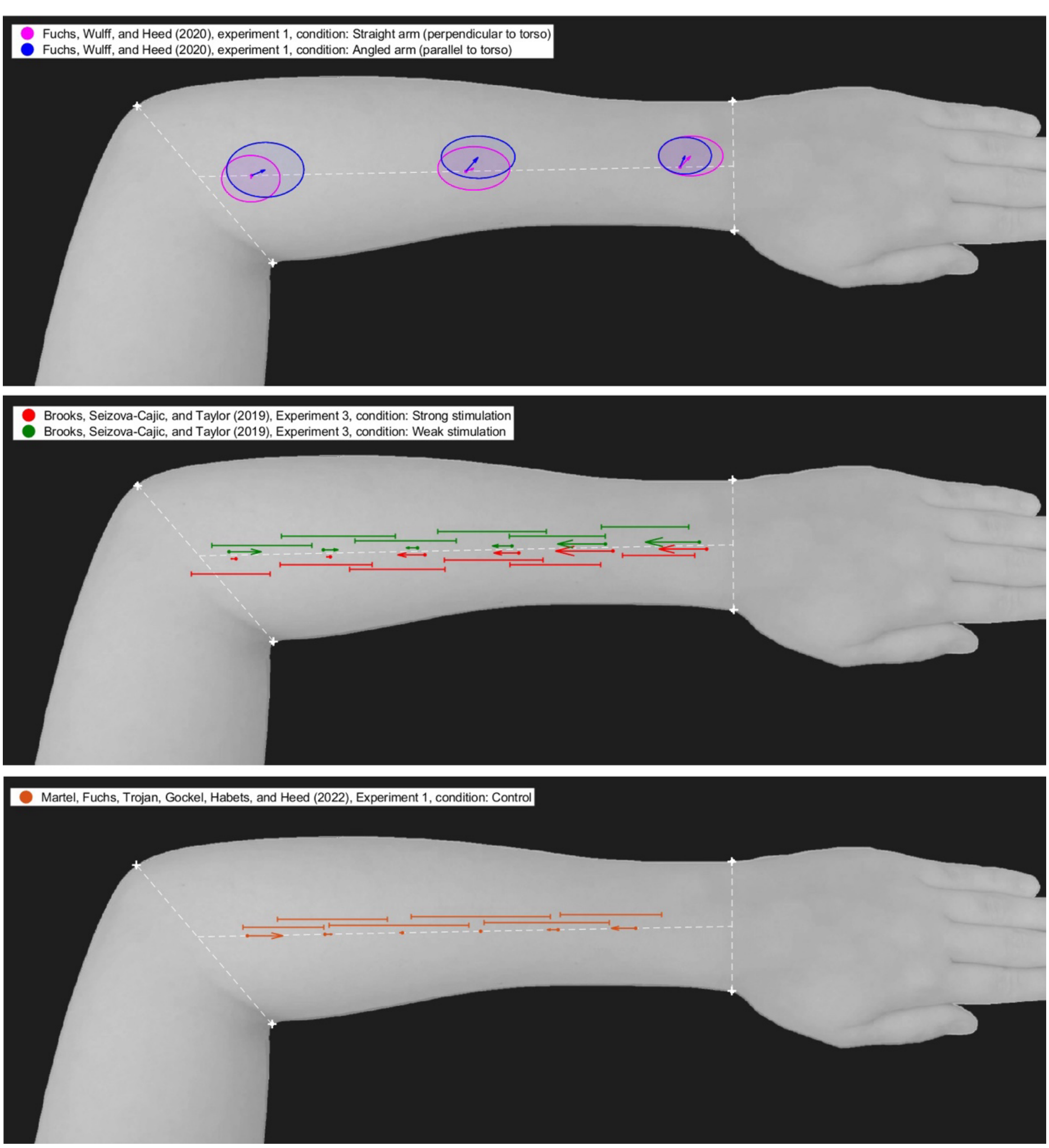

Figure 8. Response patterns for tactile localization on the forearm. Top: Data from Fuchs et al. (2020). Participants reached with their index finger and touched the stimulus location on the forearm; pink: arm perpendicular to the torso; blue: arm parallel to torso; dots: true tactile stimulus location; arrows: constant bias; ellipses: variability across participants. Middle: Data from Brooks et al. (2019), Participants indicated stimulus location with a stylus on a tablet vertically positionned next to their arm; cyan: stimulus intensity near threshold; magenta: stimulus intensity supra-threshold; variability is shown as lines because responses were assessed only in 1D along the length of the arm. Bottom: Data from Martel et al. (2022); participants indicated stimulus location with the computer mouse on an image presented on a monitor in front of them;

Fig. 8 depicts bias and precision of localization responses on the dorsal forearm reported by three exemplary studies that employed different localization tasks. The figure illustrates several important aspects.

### Bias

First, bias differed markedly between the three studies, emphasizing how the method of choice may affect seemingly simple experimental measures. The study from the top panel shows a distal bias (i.e., towards the wrist), while the studies from the middle and bottom panels indicate a repulsion effect near landmarks. Second, both body posture (top panel) and stimulus intensity (middle panel) affected localization error, suggesting that even small details of the setup may affect localization. Notably, it is not clear whether the posture bias (top panel) stems from posture affecting tactile-perceptual location or motor performance; similarly, it is not clear which aspects of each study are responsible for the different error patterns – response modality (here: direct pointing, stylus, mouse), spatial transformations required to respond on the arm vs. on a horizontal tablet vs. on a vertical screen, or yet another aspect.

### Variability

Conversely to the pronounced differences observed in bias, response variability tends to be higher in the middle of the forearm across studies. Once again, the reason for this discrepancy – differences in one aspect of performance and similarities in another – remains unclear.

The inconsistencies both in terms of findings and in terms of what is reported are often overlooked; indeed, many studies tend to focus on only one dimension, either bias or variability. For example, Miller's trilateration model (2022) accounts for variability but does not extend its explanation to the observed patterns of bias. Taken together, these results suggest that patterns of tactile localization are strongly influenced by the methods used to measure them. Finally, many past publications have not reported sufficient data to create illustrations such as Fig. 8. Visualizing more previous results could be visualized side-by-side may help advance experimentation and theorizing. We invite researchers to participate and upload their past or new results to https://osf.io/ryh52/ (detailed instructions there).

# 6 General discussion

We reviewed more than 100 scientific publications on tactile localization and scrutinized the employed experimental paradigms for their implicit assumptions, biases, and limitations. Ultimately, we seek to make explicit how tactile localization is currently conceptualized and may best be systematically studied.

Our analysis revealed four key insights. First, there is a striking variability in the employed methods, with a conspicuous variability in how much participants' responses imply spatial processing or not. Second, these methodological differences are not merely procedural but evoke marked differences in spatial bias and variability across tasks. Third, most studies have investigated tactile localization on the hand and forearm, and much less data exists for other body locations. And fourth, many tasks lack ecological validity, raising questions about the generalizability of scientific findings to natural behaviors. We discuss each point below.

## 6.1 Diversity in localization methods implies lack of agreed-upon theoretical underpinnings

Many diverse tasks are currently subsumed under the umbrella term "tactile localization". While some variations in task design stem from specific theoretical underpinnings, it is often not clear what presumed cognitive functions, mechanisms, or processes are involved with the various paradigms. Moreover, it is our impression that some task aspects have been introduced out of practical necessity more than based on theory. Accordingly, some methods appear as "best compromises" rather than tightly controlled implementations of theoretical aspects.

For example, most experiments have precluded participants from touching the target location on their own body and have instead asked that pointing responses end slightly above the skin (e.g., enforced by a barrier above the skin). The underlying idea is that the measured response indicates "pure" localization of the applied stimulus and is not improved by integrating new tactile feedback produced by the reaching effector touching the target area. However, such strategies necessarily introduce a mismatch between how a participant would naturally behave and what (s)he must do in the experiment; moreover, many tasks introduce additional requirements that likely recruit additional cognitive processes. For instance, asking participants to stop a reaching movement before contacting their own arm may interfere with a regular self-touch movement; barriers may disrupt movements planned to end on the body proper and, therefore, bias the assessed responses (see section 4.1.4); and using a pointer likely recruits processes involved in tool use.

The situation is not better regarding the potential requirement of spatial transformations by the different tasks, which we have discussed at length in our task classifications. There is no agreed-upon theory about how such transformations are implemented by our cognitive system. For instance, it is not known which types of transformations, such as translation, scaling, and rotation, evoke distinct cognitive processes in the context of deriving a movement plan based on tactile input. Thus, when tasks differ regarding such requirements, it is unclear whether they test comparable capacities or not. Finally, even if researchers assume that participants solve a task in a certain way, there are often alternative strategies participants may use instead. Besides introducing conceptual unclarity and uncertainty, this may lead to participant subgroups with distinct strategies, unbeknownst to the researchers.

As a result of the theoretically muddy methodological diversity, there is currently no unified, agreed-upon concept of ‘tactile localization’. In turn, it underscores the need for careful task design and interpretation, both to allow comparison across studies and to ascertain that participants’ strategies align with the experimenter’s assumptions.

## 6.2 Different localization methods lead to different results

Given that tasks differ along systematic characteristics, it may not be surprising that we observed some systematic biases between task classes. For example, studies targeting the hand have consistently revealed shifts of reported location toward the fingers when participants reported locations on a drawing or image. In contrast, large biases but inconsistencies between studies were observed when participants pointed onto a surface above the hand.

Thus, image-based methods appear to evoke a distortion related to the specific task requirements rather than to tactile localization per se. Alternatively, differences in bias may be indicative of tactile localization relying on different cognitive and/or neural processes (cf Longo et al., 2015). Again, it is evident how theoretical assumptions will guide the interpretation of experimental results.

An unclear picture emerged for experiments focused on the arm. Some studies reported consistent biases, such as repulsion effects near landmarks like the wrist or elbow (Brooks et al., 2019; Martel et al., 2022), while others did not (Fuchs et al., 2020). However, one observation that was consistent across the three analyzed studies was an inverted U-shaped pattern of variability: response variability tended to be higher in the middle of the arm and lower near landmarks. This pattern is consistent with the hypothesis that the somatosensory system uses a multilateration computation and thus with theoretical models, such as the Bayesian formulation of multilateration, which states that localization accuracy decreases with distance from known reference points (Cholewiak & Collins, 2003; Miller et al., 2022).

In sum, our observations here are a reminder that even seemingly simple and small details can strongly affect experimental outcomes and the derived interpretation, and prompt careful inspection of existing designs for future research.

## 6.3 Theoretical frameworks: current status and future directions

Overall, studies to date were not always clearly derived from theory. Still, several theoretical frameworks exist, and each of them has stimulated experimental work. However, the frameworks appear rather independent and, maybe more importantly, differ in the breadth of phenomena they attempt to explain. Representational models (Longo et al., 2010; Tamè et al., 2019) have tried to structure the representations involved in research that includes tactile localization but is overall larger, asking how body, skin, and posture are processed across sensory modalities. The remaining frameworks we have covered here have picked out more focused aspects. The various frameworks are partly compatible and could potentially be integrated. However, some frameworks appear contradictory in that they offer incompatible explanations for a given phenomenon. It may be at those intersections that new work derived from the different previous frameworks could be most fruitful.

It would, for instance, seem natural to integrate sensorimotor contingency with Bayesian approaches, with priors emerging from experience of contingencies throughout development and training. Such an overarching approach might attempt to specify what types of priors exist and how they are neurally coded, integrated, and weighted. Aspects of remapping – whether recoding into specific spatial formats occurs mandatorily or not, and under which conditions – would also seem to integrate well with such a framework.

The remapping account could also be conceptually integrated in the various representational models, if one assumes that the brain entertains multiple spatial maps, an idea that has been prominent in sensorimotor neuroscience (e.g., Pouget et al., 2002). Critically, the very different conclusions drawn about localization errors when assuming that they represent maps vs. an integrative process could stimulate work that would revise a representation-centered view into a view that explicitly comprises the transformations performed between specified representational formats. Attempts to reconcile some of the previous experimental approaches might, for instance, lead to removal of some previously proposed maps in favor of a common integrative mechanism that integrates information from a smaller number or maps depending on task context.

## 6.4 Focus on hand and arm and ecological validity

There is a curious dichotomy regarding ecological validity in tactile localization research. On one side, most work has investigated the hands and forearms. One might contend that this is adequate because the forelimbs are paramount to human skillful action competence. Yet, this focus on the forelimbs implies neglect of the remaining body. This lack of experimental work has implications for understanding whether tactile processing is consistent across the body. Speculatively, for example, touch may be localized using specific strategies on body parts that are always (e.g., the back) or often (e.g., the legs) out of view. Similarly, tactile priors may be much coarser for body regions that are typically not directly involved in goal-directed actions. Similar to the choice of tasks discussed above, the focus on forearms and hands in localization research probably rather reflects pragmatic choices than theoretical reasoning. Localization research on the arms and hands is relatively straightforward because the body parts can be positioned and reached easily, which facilitates stimulation and localization procedures, and there are plenty of reference studies that can serve as templates.

As a related point, the choice of experimental methods is often far from what we encounter in everyday life. For example, tasks often isolate tactile localization from its natural context, where it is embedded in dynamic, multisensory, and sensorimotor interactions. Natural tactile localization often occurs as a closed feedback loop, that is, a tactile-sensory input evokes a motor response which, in turn, creates additional input that allows evaluating the action with respect to the initial tactile input. For instance, when one scratches an itch, the scratching creates new tactile as well as proprioceptive input, and one can determine whether one has scratched the correct location and, if not, move the scratching hand along until the itch is found. Thus, real-life action relies on continuous sensory feedback and is inherently multisensory. In contrast, most experimental paradigms artificially separate the target body part from the effector and restrict the use of non-tactile information.

Much has been said in recent years about the potential pitfalls of experimental tasks that do not line up with naturalistic behaviors (e.g., Krakauer et al., 2017). The aim of highly controlled paradigms has been to isolate specific mechanisms, but the reductionistic methodology likely creates findings that often do not generalize to naturalistic settings (e.g., Cisek & Green, 2024). The counter-argument is, of course, that well-controlled experimentation allows isolating the individual factors that affect the behavior of interest. It seems to us, though, that in the case of tactile localization the intended precision afforded by experimental tasks is often degraded by the implicit use of multiple unspecified and untested assumptions inherent in the tasks.

## 6.5 Conclusions

Tactile localization is surprisingly complex. Much of this complexity is owed to a large variability in experimental approaches and, partly, in (sometimes lack of) the respective theoretical underpinnings. Whereas there is a scattered picture of experimental findings as well as theoretical approaches, there are obvious opportunities for integrating currently rather separate theories and frameworks. In part, such scientific convergence would require careful and explicit consideration of the cognitive requirements induced by each experimental task. Such scrutiny will, in turn, allow systematic exploration of task and other context variables. As is currently the case in many areas of cognitive neuroscience, there is increasing awareness of the distance between well-controlled experimental methodology and its validity with respect to natural behavior. Developing naturalistic tasks while retaining sufficient experimental control for theoretical rigor may be an important task for research to come. Finally, our comparisons of original data highlights that making experimental data available could support also larger-scale comparisons, which could further illuminate common trends across the various facets of tactile localization. We look forward to everyone’s contribution.

# 7 Declarations

## 7.1 Funding

X.F. was supported by the University Salzburg Early-Career Grant “SOMASTI” (IPE184001_03) and a fellowship by the Aktion Österreich-Tschechien (MSMT-2677/2023-6).
J.A.M.K., S.T.P., and M.H. were supported by the Czech Science Foundation (GA ČR), project no. 25-18113S.
S.T.P. was additionally supported by the project Mobility ČVUT MSCA-F-CZ-I, no. CZ.02.01.01/00/22_010/0003405 by the Ministry of Education, Youth, and Sports, Czech Republic.

## 7.2 Conflicts of interest

The authors declare that there were no conflicts of interest.

## 7.3 Ethics approval

Not applicable (no tested participants)

## 7.4 Consent to participate

Not applicable (no tested participants)

## 7.5 Consent for publication

Not applicable (no tested participants)

## 7.6 Availability of data and materials

All material is available at https://osf.io/ryh52/, including the data extracted from the literature that we used for the analyses.

## 7.7 Code availability

All code is available at https://osf.io/ryh52/.

## 7.8 Authors' contributions

Author contributions statements as stipulated in the *Contributor Roles Taxonomy* (CRediT):

- Xaver Fuchs: Conceptualization; Data curation; Formal analysis; Funding acquisition; Investigation; Methodology; Project administration; Resources; Supervision; Validation; Visualization; Writing original draft; Writing review and editing.
- Jason A. M. Khoury: Conceptualization; Data curation; Formal analysis; Investigation; Methodology; Project administration; Supervision; Validation; Visualization; Writing original draft; Writing review and editing.
- Sergiu Tcaci Popescu: Conceptualization; Data curation; Funding acquisition; Investigation; Methodology; Project administration; Supervision; Validation; Writing review and editing.
- Tobias Heed: Conceptualization; Project administration; Supervision; Validation; Writing review and editing.
- Matej Hoffmann: Conceptualization; Funding acquisition; Project administration; Resources; Supervision; Validation; Writing review and editing.

## 7.9 Acknowledgements

We would like to thank Souta Hidaka, Matthew Longo and Jared Medina for providing us data for our quantitative comparison between localization methods.

# The Where and How of Touch—Supplementary Material

# 1 Included studies

In total, 106 studies were included in the review (Albanese et al., 2009; Antfolk et al., 2010; Auclair et al., 2012; Azanon et al., 2010; Bara-Jimenez, Shelton, & Hallett, 2000; Bara-Jimenez, Shelton, Sanger, et al., 2000; Benedetti, 1988; Bobich et al., 2007; Brandes & Heed, 2015; Braun et al., 2011; Braun, Hess, et al., 2005; Braun, Ladda, et al., 2005; Brooks et al., 2019; Byl et al., 1996; Cesini et al., 2020; Cholewiak et al., 2004; Cholewiak & Collins, 2003; Cholewiak & McGrath, 2006; Cicmil et al., 2016; Cody et al., 2008, 2010; Costantini et al., 2020; Craig & Rhodes, 1992; Culver, 1970; Diener et al., 2017; Dobrzynski et al., 2012; Eguchi et al., 2022; Elithorn et al., 1953; Ferre et al., 2013; Fuchs et al., 2020; Geffen et al., 1985; Gherri et al., 2022; Gilliland & Schlegel, 1994; Halnan & Wright, 1960; Harrar et al., 2013; Harrar & Harris, 2009, 2010; Harris et al., 2004, 2006; Harvie et al., 2017; Hense et al., 2019; Hidaka et al., 2020; Higashiyama & Tashiro, 1990; Ho & Spence, 2007; Hsu & Nelson, 1981; Isakovic et al., 2022; Jobst et al., 1997; Jouybari et al., 2021; Kang & Longo, 2023; Lambercy et al., 2013; Liu & Medina, 2021; Longo et al., 2015; Longo & Morcom, 2016; Macklin et al., 2021; Mancini et al., 2011; Manser-Smith et al., 2018, 2019; Margolis & Longo, 2015; Marshall, 1956; Martel et al., 2022; Martikainen & Pertovaara, 2002; Mattioni & Longo, 2014; J. Medina & Duckett, 2017; J. Medina & Rapp, 2014; S. Medina et al., 2018; Merz et al., 2020; Mikula et al., 2018; Miller et al., 2022; Moore & Schady, 1995; Nakada, 1993; Nguyen et al., 2020; Overvliet et al., 2011; Paillard et al., 1983; Pritchett et al., 2012; Pritchett & Harris, 2011; Richardson et al., 2022; Sadibolova, Tame, & Longo, 2018; Sadibolova, Tame, Walsh, et al., 2018; Samad & Shams, 2016, 2018; Schlereth et al., 2001; Seizova-Cajic & Taylor, 2014; Seyal et al., 1997; Sherrick et al., 1990; Steenbergen et al., 2012, 2013, 2014; Stevens & Choo, 1996; Stevens & Pattersen, 1995; Stopford, 1921; Tommerdahl et al., 2007; Trojan et al., 2010, 2014, 2019; J. van Erp, 2005; J. B. F. van Erp, 2008; Verhaar et al., 2022; Wahn & Koenig, 2015; Wang et al., 2018; Willis et al., 2019, 2021; Wilson et al., 2014; Yeganeh, Makarov, Stefansson Thors, et al., 2023; Yeganeh, Makarov, Unnthorsson, et al., 2023; Ylioja et al., 2006; Yoshioka et al., 2013).

# 2 Categories of localization tasks

Here we give an overview which tasks we assigned to the categories (A-I) and which tasks did not fall into any of these categories. All tasks are listed in Table S2.

| First author | Year | Loc. Resp. spatial | Spatial transf. | Loc task req. | Loc. reference | Loc. Reporting method | Task category |
|---|---|---|---|---|---|---|---|
| Bara-Jimenez | 2000a | spatial | none | indicate | absolute | reach/point | A |
| Brandes | 2015 | spatial | none | indicate | absolute | reach/point | A |
| Brandes | 2015 | spatial | none | indicate | absolute | reach/point | A |
| Byl | 1996 | spatial | none | indicate | absolute | reach/point | A |
| Ferre | 2013 | spatial | none | indicate | absolute | reach/point | A |
| Ferre | 2013 | spatial | none | indicate | absolute | reach/point | A |
| Fuchs | 2020 | spatial | none | indicate | absolute | reach/point | A |
| Fuchs | 2020 | spatial | none | indicate | absolute | reach/point | A |
| Fuchs | 2020 | spatial | none | indicate | absolute | reach/point | A |
| Fuchs | 2020 | spatial | none | indicate | absolute | reach/point | A |
| Geffen | 1985 | spatial | none | indicate | absolute | reach/point | A |
| Geffen | 1985 | spatial | none | indicate | absolute | reach/point | A |
| Harvie | 2017 | spatial | none | indicate | absolute | reach/point | A |
| Harvie | 2017 | spatial | none | indicate | absolute | reach/point | A |
| Higashiyama | 1990 | spatial | none | indicate | absolute | reach/point | A |
| Higashiyama | 1990 | spatial | none | indicate | absolute | reach/point | A |
| Higashiyama | 1990 | spatial | none | indicate | absolute | reach/point | A |
| Higashiyama | 1990 | spatial | none | indicate | absolute | reach/point | A |
| Higashiyama | 1990 | spatial | none | indicate | absolute | reach/point | A |
| Jobst | 1997 | spatial | none | indicate | absolute | reach/point | A |
| Longo | 2015 | spatial | none | indicate | absolute | reach/point | A |
| Martikainen | 2002 | spatial | none | indicate | absolute | reach/point | A |
| Medina | 2014 | spatial | none | indicate | absolute | reach/point | A |
| Mikula | 2018 | spatial | none | indicate | absolute | reach/point | A |
| Moore | 1995 | spatial | none | indicate | absolute | reach/point | A |
| Nakada | 1993 | spatial | none | indicate | absolute | reach/point | A |
| Paillard | 1983 | spatial | none | indicate | absolute | reach/point | A |
| Seizova-Cajic | 2014 | spatial | none | indicate | absolute | reach/point | A |
| Trojan | 2014 | spatial | none | indicate | absolute | reach/point | A |
| Trojan | 2019 | spatial | none | indicate | absolute | reach/point | A |
| Yoshioka | 2013 | spatial | none | indicate | absolute | reach/point | A |
| Brooks | 2019 | spatial | transl. only | indicate | absolute | reach/point | B or C |
| Brooks | 2019 | spatial | transl. only | indicate | absolute | reach/point | B or C |
| Brooks | 2019 | spatial | transl. only | indicate | absolute | reach/point | B or C |
| Fuchs | 2020 | spatial | transl. only | indicate | absolute | reach/point | B or C |
| Fuchs | 2020 | spatial | transl. only | indicate | absolute | reach/point | B or C |
| Fuchs | 2020 | spatial | transl. only | indicate | absolute | reach/point | B or C |
| Harrar | 2010 | spatial | transl. only | indicate | absolute | reach/point | B or C |
| Hidaka | 2020 | spatial | transl. only | indicate | absolute | reach/point | B or C |
| Longo | 2015 | spatial | transl. only | indicate | absolute | reach/point | B or C |
| Longo | 2016 | spatial | transl. only | indicate | absolute | reach/point | B or C |
| Mattioni | 2014 | spatial | transl. only | indicate | absolute | reach/point | B or C |
| Medina | 2017 | spatial | transl. only | indicate | absolute | reach/point | B or C |
| Medina | 2017 | spatial | transl. only | indicate | absolute | reach/point | B or C |
| Medina | 2017 | spatial | transl. only | indicate | absolute | reach/point | B or C |
| Steenbergen | 2012 | spatial | transl. only | indicate | absolute | reach/point | B or C |
| Steenbergen | 2013 | spatial | transl. only | indicate | absolute | reach/point | B or C |
| Steenbergen | 2014 | spatial | transl. only | indicate | absolute | reach/point | B or C |
| Stopford | 1921 | spatial | transl. only | indicate | absolute | reach/point | B or C |
| Stopford | 1921 | spatial | transl. only | indicate | absolute | reach/point | B or C |
| Trojan | 2010 | spatial | transl. only | indicate | absolute | reach/point | B or C |
| Trojan | 2010 | spatial | transl. only | indicate | absolute | reach/point | B or C |
| Craig | 1992 | spatial | multiple | indicate | absolute | show on image | D, E, or G |
| Culver | 1970 | spatial | multiple | indicate | absolute | show on image | D, E, or G |
| Elithorn | 1953 | spatial | multiple | indicate | absolute | show on image | D, E, or G |
| Halnan | 1960 | spatial | multiple | indicate | absolute | show on image | D, E, or G |
| Ho | 2007 | spatial | multiple | indicate | absolute | show on image | D, E, or G |
| Kang | 2023 | spatial | multiple | indicate | absolute | show on image | D, E, or G |
| Macklin | 2021 | spatial | multiple | indicate | absolute | show on image | D, E, or G |
| Mancini | 2011 | spatial | multiple | indicate | absolute | show on image | D, E, or G |
| Mancini | 2011 | spatial | multiple | indicate | absolute | show on image | D, E, or G |
| Mancini | 2011 | spatial | multiple | indicate | absolute | show on image | D, E, or G |
| Margolis | 2015 | spatial | multiple | indicate | absolute | show on image | D, E, or G |
| Martel | 2022 | spatial | multiple | indicate | absolute | show on image | D, E, or G |
| Medina | 2018 | spatial | multiple | indicate | absolute | show on image | D, E, or G |
| Merz | 2020 | spatial | multiple | indicate | absolute | show on image | D, E, or G |
| Miller | 2021 | spatial | multiple | indicate | absolute | show on image | D, E, or G |
| Miller | 2021 | spatial | multiple | indicate | absolute | show on image | D, E, or G |
| Pritchett | 2012 | spatial | multiple | indicate | absolute | show on image | D, E, or G |
| Pritchett | 2012 | spatial | multiple | indicate | absolute | show on image | D, E, or G |
| Sadibolova | 2018a | spatial | multiple | indicate | absolute | show on image | D, E, or G |
| Cholewiak | 2003 | spatial | transl. only | decide | absolute | button press | F |
| Cholewiak | 2003 | spatial | transl. only | decide | absolute | button press | F |
| Cholewiak | 2003 | spatial | transl. only | decide | absolute | button press | F |
| Cholewiak | 2003 | spatial | transl. only | decide | absolute | button press | F |
| Verhaar | 2022 | spatial | transl. only | decide | absolute | button press | F |
| Antfolk | 2010 | spatial | multiple | decide | absolute | verbal | H |
| Bobich | 2007 | spatial | multiple | decide | absolute | verbal | H |
| Braun | 2005a | spatial | multiple | decide | absolute | verbal | H |
| Cesini | 2020 | spatial | multiple | decide | absolute | verbal | H |
| Marshall | 1956 | spatial | multiple | decide | absolute | verbal | H |
| Auclair | 2012 | non-spatial | none | decide | absolute | verbal | I |
| Auclair | 2012 | non-spatial | none | decide | absolute | verbal | I |
| Auclair | 2012 | non-spatial | none | decide | absolute | verbal | I |
| Auclair | 2012 | non-spatial | none | decide | absolute | verbal | I |

| Auclair | 2012 | non-spatial | none | decide | absolute | verbal | I |
|---|---|---|---|---|---|---|---|
| Cicmil | 2016 | non-spatial | none | decide | absolute | verbal | I |
| Halnan | 1960 | non-spatial | none | decide | absolute | verbal | I |
| Halnan | 1960 | non-spatial | none | decide | absolute | verbal | I |
| Harris | 2004 | non-spatial | none | decide | absolute | verbal | I |
| Harris | 2004 | non-spatial | none | decide | absolute | verbal | I |
| Harris | 2004 | non-spatial | none | decide | absolute | verbal | I |
| Harris | 2004 | non-spatial | none | decide | absolute | verbal | I |
| Manser-Smith | 2018 | non-spatial | none | decide | absolute | verbal | I |
| Manser-Smith | 2018 | non-spatial | none | decide | absolute | verbal | I |
| Manser-Smith | 2019 | non-spatial | none | decide | absolute | verbal | I |
| Manser-Smith | 2019 | non-spatial | none | decide | absolute | verbal | I |
| Stevens | 1995 | non-spatial | none | decide | absolute | verbal | I |
| Yeganeh | 2023b | non-spatial | none | decide | relative | button press | J |
| Albanese | 2009 | non-spatial | none | decide | absolute | button press | none |
| Azanon | 2010 | spatial | none | decide | relative | verbal | none |
| Bara-Jimenez | 2000b | spatial | multiple | decide | absolute | show on image | none |
| Benedetti | 1988 | spatial | multiple | other | other | other | none |
| Braun | 2005b | spatial | multiple | decide | absolute | show on image | none |
| Braun | 2011 | non-spatial | none | decide | absolute | button press | none |
| Braun | 2011 | non-spatial | none | decide | absolute | button press | none |
| Byl | 1996 | non-spatial | none | decide | absolute | movement | none |
| Cholewiak | 2004 | spatial | none | decide | absolute | button press | none |
| Cholewiak | 2004 | spatial | multiple | decide | absolute | button press | none |
| Cholewiak | 2004 | spatial | multiple | decide | absolute | button press | none |
| Cholewiak | 2006 | spatial | multiple | decide | absolute | button press | none |
| Cholewiak | 2006 | spatial | multiple | decide | absolute | button press | none |
| Cholewiak | 2006 | spatial | multiple | decide | absolute | button press | none |
| Cholewiak | 2006 | spatial | multiple | decide | absolute | button press | none |
| Cody | 2008 | spatial | none | decide | relative | verbal | none |
| Cody | 2010 | spatial | none | decide | relative | verbal | none |
| Costantini | 2020 | spatial | multiple | decide | absolute | show on image | none |
| Diener | 2017 | spatial | multiple | decide | absolute | show on image | none |
| Dobrzynski | 2011 | spatial | multiple | decide | absolute | show on image | none |
| Eguchi | 2022 | spatial | transl. only | indicate | absolute | other | none |
| Gherri | 2022 | spatial | multiple | decide | absolute | button press | none |
| Gilliland | 1994 | spatial | multiple | decide | absolute | button press | none |
| Gilliland | 1994 | spatial | multiple | decide | absolute | button press | none |
| Gilliland | 1994 | spatial | multiple | decide | absolute | button press | none |
| Harrar | 2009 | spatial | transl. only | indicate | absolute | verbal | none |
| Harrar | 2009 | spatial | transl. only | indicate | absolute | verbal | none |
| Harrar | 2009 | spatial | transl. only | indicate | absolute | verbal | none |
| Harrar | 2013 | spatial | none | decide | absolute | verbal | none |
| Harris | 2006 | spatial | none | decide | absolute | reach/point | none |
| Harris | 2006 | non-spatial | none | decide | absolute | movement | none |
| Harris | 2006 | spatial | none | decide | absolute | reach/point | none |
| Harris | 2006 | non-spatial | none | decide | absolute | movement | none |
| Harvie | 2017 | spatial | multiple | decide | absolute | button press | none |
| Harvie | 2017 | spatial | multiple | decide | absolute | button press | none |
| Hense | 2019 | spatial | multiple | decide | absolute | movement | none |
| Higashiyama | 1990 | spatial | none | decide | absolute | verbal | none |
| Higashiyama | 1990 | spatial | none | decide | absolute | verbal | none |
| Higashiyama | 1990 | spatial | none | decide | absolute | verbal | none |
| Higashiyama | 1990 | spatial | none | decide | absolute | verbal | none |
| Higashiyama | 1990 | spatial | none | decide | absolute | verbal | none |
| Isakovic | 2022 | spatial | multiple | decide | absolute | show on image | none |
| Jouybari | 2021 | spatial | multiple | decide | absolute | button press | none |
| Lambercy | 2013 | spatial | transl. only | decide | absolute | show on image | none |
| Liu | 2021 | spatial | transl. only | decide | absolute | verbal | none |
| Liu | 2021 | spatial | transl. only | decide | absolute | verbal | none |
| Liu | 2021 | spatial | transl. only | decide | absolute | verbal | none |
| Longo | 2015 | spatial | transl. only | indicate | absolute | show on image | none |
| Mancini | 2011 | spatial | multiple | indicate | absolute | other | none |
| Marshall | 1956 | spatial | multiple | decide | absolute | show on image | none |
| Merz | 2020 | spatial | transl. only | decide | absolute | reach/point | none |
| Miller | 2021 | spatial | transl. only | indicate | absolute | show on image | none |
| Overvliet | 2011 | spatial | transl. only | decide | absolute | verbal | none |
| Overvliet | 2011 | spatial | transl. only | decide | absolute | verbal | none |
| Pritchett | 2011 | spatial | transl. only | decide | relative | button press | none |
| Pritchett | 2012 | spatial | transl. only | indicate | absolute | show on image | none |
| Richardson | 2022 | spatial | multiple | decide | absolute | button press | none |
| Richardson | 2022 | spatial | multiple | decide | absolute | button press | none |
| Sadibolova | 2018a | spatial | multiple | indicate | absolute | other | none |
| Sadibolova | 2018b | spatial | transl. only | indicate | absolute | show on image | none |
| Samad | 2016 | spatial | none | indicate | absolute | show on image | none |
| Samad | 2018 | spatial | none | indicate | absolute | show on image | none |
| Seyal | 1997 | spatial | none | decide | absolute | verbal | none |
| Seyal | 1997 | spatial | none | decide | absolute | verbal | none |
| Sherrick | 1990 | spatial | multiple | decide | absolute | button press | none |
| Sherrick | 1990 | spatial | multiple | decide | absolute | button press | none |
| Sherrick | 1990 | spatial | multiple | decide | absolute | button press | none |
| Sherrick | 1990 | spatial | multiple | decide | absolute | button press | none |
| Stevens | 1996 | spatial | multiple | decide | relative | verbal | none |
| Stopford | 1921 | spatial | multiple | indicate | absolute | reach/point | none |
| Tommerdahl | 2007 | non-spatial | none | decide | relative | verbal | none |
| Wahn | 2015 | spatial | transl. only | decide | relative | button press | none |
| Wang | 2018 | spatial | multiple | decide | absolute | button press | none |
| Willis | 2019 | spatial | transl. only | decide | absolute | verbal | none |
| Willis | 2021 | spatial | transl. only | decide | absolute | verbal | none |
| Wilson | 2014 | spatial | transl. only | indicate | absolute | show on image | none |
| Yeganeh | 2023a | spatial | multiple | decide | relative | button press | none |
| Yeganeh | 2023a | spatial | multiple | decide | relative | button press | none |
| van Erp | 2005 | spatial | multiple | decide | relative | button press | none |
| van Erp | 2005 | spatial | multiple | decide | relative | button press | none |
| van Erp | 2008 | spatial | transl. only | other | other | other | none |

Table S2: Tasks assigned to classification categories.

## 2.1 Category A (direct pointing)

31 tasks coming from 20 studies fell into category A (Bara-Jimenez, Shelton, & Hallett, 2000; Brandes & Heed, 2015; Byl et al., 1996; Ferre et al., 2013; Fuchs et al., 2020; Geffen et al., 1985; Harvie et al., 2017; Higashiyama & Tashiro, 1990; Jobst et al., 1997; Longo et al., 2015; Martikainen & Pertovaara, 2002; J. Medina & Rapp, 2014; Mikula et al., 2018; Moore & Schady, 1995; Nakada, 1993; Paillard et al., 1983; Seizova-Cajic & Taylor, 2014; Trojan et al., 2014, 2019; Yoshioka et al., 2013).

## 2.2 Category B and C (reaching/pointing with translation)

21 tasks coming from 13 studies fell into category B or C (Brooks et al., 2019; Fuchs et al., 2020; Harrar & Harris, 2010; Hidaka et al., 2020; Longo et al., 2015; Longo & Morcom, 2016; Mattioni & Longo, 2014; J. Medina & Duckett, 2017; Steenbergen et al., 2012, 2013, 2014; Stopford, 1921; Trojan et al., 2010).

## 2.3 Category D, E, and G (showing locations on an image)

19 tasks from 12 studies fell into either category D, E, or G (Craig & Rhodes, 1992; Culver, 1970; Elithorn et al., 1953; Halnan & Wright, 1960; Ho & Spence, 2007; Kang & Longo, 2023; Martel et al., 2022; S. Medina et al., 2018; Merz et al., 2020; Miller et al., 2022; Pritchett et al., 2012; Sadibolova, Tame, & Longo, 2018).

## 2.4 Category F (button presses with translation)

5 tasks coming from two studies fell into category F (Cholewiak & Collins, 2003; Verhaar et al., 2022).

## 2.5 Category H (verbal responses with multiple transformations)

5 tasks coming from 5 studies fell into category H (Antfolk et al., 2010; Bobich et al., 2007; Braun, Hess, et al., 2005; Cesini et al., 2020; Marshall, 1956).

## 2.6 Category I (verbally naming body parts)

17 tasks from 7 studies fell into category I (Auclair et al., 2012; Cicmil et al., 2016; Halnan & Wright, 1960; Harris et al., 2004; Manser-Smith et al., 2018, 2019; Stevens & Pattersen, 1995).

## 2.7 Category J (relative decision)

One task fell into category J (Yeganeh, Makarov, Unnthorsson, et al., 2023).

## 2.8 Tasks not falling into any of the above categories

81 tasks from 55 studies did not fall into any of the above categories (Albanese et al., 2009; Azanon et al., 2010; Bara-Jimenez, Shelton, Sanger, et al., 2000; Benedetti, 1988; Braun et al., 2011; Braun, Ladda, et al., 2005; Byl et al., 1996; Cholewiak et al., 2004; Cholewiak & McGrath, 2006; Cody et al., 2008, 2010; Diener et al., 2017; Dobrzynski et al., 2012; Eguchi et al., 2022; Gherri et al., 2022; Gilliland & Schlegel, 1994; Halnan & Wright, 1960; Harrar et al., 2013; Harrar & Harris, 2009; Harris et al., 2006; Harvie et al., 2017; Hense et al., 2019; Higashiyama & Tashiro, 1990; Isakovic et al., 2022; Jouybari et al., 2021; Lambercy et al., 2013; Liu & Medina, 2021; Longo et al., 2015; Macklin et al., 2021; Mancini et al., 2011; Margolis & Longo, 2015; Marshall, 1956; Merz et al., 2020; Miller et al., 2022; Overvliet et al., 2011; Pritchett et al., 2012; Pritchett & Harris, 2011; Richardson et al., 2022; Sadibolova, Tame, & Longo, 2018; Sadibolova, Tame, Walsh, et al., 2018; Samad & Shams, 2016, 2018; Seyal et al., 1997; Sherrick et al., 1990; Stevens & Choo, 1996; Stopford, 1921; Tommerdahl et al., 2007; J. van Erp, 2005; J. B. F. van Erp, 2008; Wahn & Koenig, 2015; Wang et al., 2018; Willis et al., 2019, 2021; Wilson et al., 2014; Yeganeh, Makarov, Stefansson Thors, et al., 2023).

# 3 Figures and Tables

## 3.1 Data extraction for Fig. 7 and Fig. 8

We identified 10 studies that corresponded to the criteria “Data Availability”, “Stimulus Location”, and “Response Type” as described in the section 5 Quantitative comparison of localization methods in the manuscript. In order to extract the results and visualize them together, we used WebPlotDigitizer (https://apps.automeris.io/wpd4/) and a custom program developed in MATLAB (version R2023b). The extracted points were then mapped onto common images (either a forearm or a hand).

### 3.1.1 Dorsal hand (Fig. 7)

We also compared the results from seven selected on the dorsal hand (Hidaka et al., 2020; Kang & Longo, 2023; Longo et al., 2015; Mancini et al., 2011; Margolis & Longo, 2015; J. Medina & Duckett, 2017; S. Medina et al., 2018).

We compared two different methods that are pointing to a board on top of the hand using a baton and showing on an image or a silhouette of the participant's own hand. For the pointing method (left panel), the raw data of the responses were kindly provided by the authors of the respective studies upon request. For the responses on the image or silhouette (right panel) the figures from each article were opened in Matlab (version 2023b) and the function *ginput* was used to (1) get both the targets and the participants mean response for each locations, (2) get the anatomical landmarks of the hand provided by the article with an example of targets displayed on the hand, (3) get the same anatomical landmark in our reference hand (i.e., the image of the hand that we used to plot the data, see Fig. 6 in the article). Following the extraction of the coordinates, a thin plate spline transformation was performed to interpolate the points from the results sections to the targets on the picture of the hand provided in the article and use the transformation matrix to interpolate the responses on that hand as well. Next, we interpolated the anatomical landmarks of the picture of the hand provided to our reference hand, and finally we used the transformation matrix to interpolate the previously transformed targets and results on our reference image.

### 3.1.2 Dorsal forearm (Fig. 8)

To compare the results from the three selected studies on the forearm (Brooks et al., 2019; Fuchs et al., 2020; Martel et al., 2022), we used the image of a dorsal forearm provided by (Martel et al., 2022).

The first step was to identify the landmarks (i.e., inner and outer wrist and elbow) that would be used as a reference to transform the data from the article onto our forearm. The landmarks were identified and located using the function *ginput* to manually select the four points, inner wrist, outer wrist, inner elbow and outer elbow. These selections were marked with visible white cross markers, ensuring clear identification on the image. Upon identifying the landmarks, two segments were created: the wrist width connecting the inner and outer wrist landmarks and the elbow width which connects the inner and outer elbow landmarks. Each segment was depicted using dashed white lines, facilitating a clear distinction of forearm width dimensions (see Fig. 7 in the article). To be able to plot the proximo distal errors, midpoints of both the elbow and wrist landmarks were computed to plot the wrist-elbow segment. The shape of the elbow midpoint was corrected and determined to be slightly above the elbow midpoint to better represent the proximal-distal orientation of the forearm. Finally, to ensure that the results of each article were properly aligned with the image of the forearm that we used (in pixels), we computed a conversion factor using the known length (elbow to wrist) of the real arm (215 mm) provided by Martel et al. (2022).

The study by Fuchs et al. (2020) investigated tactile localization with participants having to reach/point to where they felt the tactile stimulation. Here we focused on the results of their experiment 1 (both conditions: with straight arm and angled arm) on the pointing before the correction of the localization on the skin. They used three locations on the dorsal forearm;

proximal that is close to the elbow, in the middle, and distal that is close to the wrist. The authors provided us with the data (both the constant error and the variable error). The mean forearm length among participants in that study was 219 mm. The constant errors and variable errors in the proximodistal axis were first multiplied by the mean forearm length provided then divided by the conversion factor. A similar computation was done for the constant and variable errors in the mediolateral axis.

The study by Brooks et al. (2019) investigated the effect of weak and strong mechanical touch stimuli on tactile localization bias on the dorsal forearm, participants had to report their responses on a tablet vertically positioned next to the arm. We specifically focused on the results from experiment 3. Similar to the other experiments, the forearm was obscured from participants' view by the vertically positioned graphics tablet set parallel to the forearm, but it provided a larger number of locations (compared with the other experiments) allowing us to have a better display of the results. The data were extracted directly from the article (Brooks et al. (2019), Fig. 4 A and Fig. 5) using the web application *WebPlotDigitizer*. The target locations were computed using the information provided by the article: "*six locations, each 4 cm apart, were marked on the left forearm, with the most proximal location ~3 cm from the elbow and the most distal ~3 cm from the wrist*". Hence, the locations of the targets were corrected according to this description. From the figure we could extract both the responses and the target location along the forearm (given that we knew the distances from the landmarks and between targets). The means for the responses in the same condition and location was computed. The values were normalized according to the previously computed Conversion Factor. In that study, because the instructions specified that the mediolateral axis did not matter, only the proximodistal bias was reported in the study, and thus here as well.

The study by Martel et al. (2022) investigated the cutaneous rabbit illusion. However, the authors provided us with the control experiment data where participants had to localize the tactile stimulus they received on an image of the forearm on a screen in front of them. Because the participants reported the response on the exact same image that we are using, we could report the responses of the participants without further transformation. However, the target mediolateral axis (y-axis) was not accurately recorded, as it was not essential for the primary objective of the control. Therefore, we chose to display constant and variable errors only on the proximodistal axis.